\documentclass[a4paper,fleqn,usenatbib]{mnras}

\usepackage[T1]{fontenc}
\usepackage{ae,aecompl}

\usepackage{graphicx}	
\usepackage{amsmath}	
\usepackage{amssymb}	

\title[Linear polarization of ultra-cool dwarfs]{Optical and near-infrared linear polarization of low and intermediate-gravity ultra-cool dwarfs}
\author[P. A. Miles-P\'aez et al.]{
P. A. Miles-P\'aez,$^{1,2, 6}$\thanks{E-mail: ppaez@uwo.ca}
M. R. Zapatero Osorio,$^{3}$
E. Pall\'e$^{1,2}$
and K. Pe\~na Ram\'\i rez$^{4,5}$
\\
$^{1}$Instituto de Astrof\'isica de Canarias, Calle V\'ia L\'actea s/n, 38205 La Laguna, Tenerife, Spain\\
$^{2}$Dpt. de Astrof\'isica, Univ. de La Laguna, Avda. Astrof\'isico Francisco S\'anchez s$/$n, 38206 La Laguna, Tenerife, Spain\\
$^{3}$Centro de Astrobiolog\'ia (CSIC-INTA), Carretera de Ajalvir km 4, 28850 Torrej\'on de Ardoz, Madrid, Spain\\
$^{4}$Unidad de Astronom\'ia, Facultad de Cs. B\'asicas, Univ. de Antofagasta, Av. U. de Antofagasta 02800, Antofagasta, Chile\\
$^{5}$Millennium Institute of Astrophysics, Santiago, Chile\\
$^{6}$The University of Western Ontario, Department of Physics and Astronomy, 1151 Richmond Avenue, London, ON N6A 3K7, Canada
}

\date{}

\pubyear{2016}

\begin{document}
\label{firstpage}
\pagerange{\pageref{firstpage}--\pageref{lastpage}}
\maketitle

\begin{abstract}
We aim to study the optical and near-infrared linear polarimetric properties of a sample of young M7-L7 dwarfs ($\approx\,$1-500 Myr) with spectroscopic signatures of low- and intermediate-gravity atmospheres. We collected optical ($RIZ$) and near-infrared ($YJHK_{\rm s}$) linear polarimetry images on various time scales from $\sim$0.2 h to months. Linear polarization degrees in the interval 0-1.5\% ($I$- and $J$-bands) were measured with accuracies ranging from $\pm$0.1\,\% to $\pm$0.9\,\% depending on the observing filter and the target brightness. We found that the young field dwarfs in our sample show similar polarimetric degrees at both $I$- and $J$-bands, and that there is no obvious trend with the spectral type. The two Taurus sources in our sample show intense levels of $J$-band linear polarization probably due to surrounding disks. By compiling data from the literature for high-gravity M7-L7 dwarfs with likely ages $\ge$500 Myr, we did not observe any apparent difference in the linear polarimetry intensity between the young and old samples that could be ascribed to differing atmospheric gravities. Polarimetric variability with peak to peak amplitudes up to 1.5\,\% is detected on scales of about a rotation in two out of four targets that were monitored over several hours. Long-term polarimetric variability is also detected in nearly all dwarfs of the sample with data spanning months to years.
\end{abstract}

\begin{keywords}
polarization -- brown dwarfs -- stars: atmospheres -- stars: late-type -- stars: low-mass
\end{keywords}



\section{Introduction}\label{intro}

Young L dwarfs with ages below a few hundred Myr have low to moderate surface gravitiy atmospheres \citep[log\,$g \lesssim 4.5$ cm\,s$^{-2}$;][]{2000ApJ...542..464C,2003A&A...402..701B}. This produces some effects on the observed photometric and spectroscopic properties, e.g., likely thick clouds of ``dust" leading to very red colors (low atmospheric pressures are believed to retard the precipitation of solid and liquid condensates, \citealt{2010ApJ...715..561A}), and atomic and molecular spectroscopic features of different intensity as compared to high-gravity dwarfs (e.g., \citealt{1999ApJ...519..802K,1999AJ....118.2466M,2001MNRAS.326..695L,2009AJ....137.3345C, 2013ApJ...772...79A,2014A&A...562A.127B}).

The presence of dust in the atmospheres may produce linear polarization in the emergent optical and near-infrared fluxes by means of scattering processes \citep{2001ApJ...561L.123S}. According to available models that take into account the single- and multi-scattering scenario, the net degree of linear polarization depends on how the polarimetric signals from different surface areas cancel each other. In this regard, asymmetries, e.g., a non-spherical shape and$/$or an heterogeneous distribution of the dust, may lead to levels of linear polarization of $\lesssim2$\,\%~at optical and near-infrared wavelengths \citep{2010ApJ...722L.142S,2011MNRAS.417.2874M,2011ApJ...741...59D}. This prediction agrees with the polarimetric measurements in the $I$-, $Z$-, $J$-, and $H$-filters published by \citet{2002A&A...396L..35M}, \citet{2005ApJ...621..445Z}, \citet{2009A&A...502..929G}, \citet{2009A&A...508.1423T}, \citet{2011ApJ...740....4Z}, and \citet{2013A&A...556A.125M}. The latter authors tried to relate the linear polarimetry data with the rotation rate of a limited sample of field late-M and L-type dwarfs, most of which are likely $\ge500$ Myr old and have high spectroscopic rotational velocities ($v$\,sin\,$i\ge30$\,km\,s$^{-1}$). They found typical $J$-band linear polarization degrees in the interval $\le0.4-0.8\%$, and discussed that the fastest rotating dwarfs ($v$\,sin\,$i\ge60$\,km\,s$^{-1}$) tend to have a larger fraction of positive polarimetric detections and a slightly higher averaged linear polarization degree than the objects with slower rotations. This might be explained by a different level of oblateness induced by the various rotational speeds.

Young brown dwarfs of spectral type late-M and L undergo a rotational spin-up because of gravitational contraction and many show $v$\,sin\,$i$ of a few to several tens km\,s$^{-1}$ at intermediate ages (\citealt{2006ApJ...647.1405Z}, and references therein). The low superficial gravity combined with a moderate-to-high rotation favor oblate atmospheres containing significant amounts of dust in the upper layers. Under this circumstance and according to the scattering scenario, non-zero linear polarization is thus expected. Here, we present multi-epoch, optical ($RIZ$-band) and near-infrared ($YJHK_{\rm s}$) imaging linear polarimetry of a sample of 14 M7--L7 ultra-cool dwarfs. All objects display clear spectroscopic signposts of low and intermediate gravity atmospheres. Also, many of them have quite red colors. By combining these data with the polarimetric measurements from the literature, we aim at studying whether there is a relation between linear polarization and atmospheric gravity (or age).

\begin{table*}
\caption{List of targets.}
\label{Table1}
\tiny
\centering
\renewcommand{\arraystretch}{1.0}
\setlength{\tabcolsep}{2.5pt}
\begin{tabular}{l c c r r r r c r r r l}
\hline\hline
Name	&SpT	&$J$		&\multicolumn{1}{c}{$W1$}	&\multicolumn{1}{c}{$W2$}	&\multicolumn{1}{c}{$W3$}	&\multicolumn{1}{c}{$W4$}	&\multicolumn{1}{c}{Opt/NIR}	&\multicolumn{1}{c}{Age$^{\rm a}$} &\multicolumn{1}{c}{Age$^{\rm b}$} &\multicolumn{1}{c}{Mass}    &\multicolumn{1}{l}{Ref}\\
		&		&(mag)	&\multicolumn{1}{c}{(mag)}	&\multicolumn{1}{c}{(mag)}	&\multicolumn{1}{c}{(mag)}	&\multicolumn{1}{c}{(mag)}	&  		&(Myr)  	&(Myr)  	&(M$_{\rm Jup}$)	&\\
\hline
EROS-MP J0032$-$4405           &L0	&14.78\,$\pm$\,0.04	&12.84\,$\pm$\,0.02	&12.52\,$\pm$\,0.02	&11.80\,$\pm$\,0.25	&--	                &y,y  &          	& 12--22  & $\sim$10  &1,4,13\\
2MASS J00332386$-$1521309	&L4	&15.29\,$\pm$\,0.06	&12.82\,$\pm$\,0.02	&12.50\,$\pm$\,0.03	&11.85\,$\pm$\,0.31	&--	                &y,y  & $\ge$500        & 30--50  & $\ge$60  &1,4,6\\
2MASS J00452143$+$1634446 	&L2	&13.06\,$\pm$\,0.02	&10.78\,$\pm$\,0.02	&10.40\,$\pm$\,0.02	& 9.80\,$\pm$\,0.05	&8.64\,$\pm$\,0.34	&y,y  & 10$-$100	& 30--50  & 12--25   &1,4,6\\
2MASS J02411151$-$0326587 	&L0	&15.80\,$\pm$\,0.07	&13.65\,$\pm$\,0.03	&13.28\,$\pm$\,0.03	&--	                &--	                &y,y  & $\ge$500	& 20--40  & $\ge60$  &1,4,6\\
2MASS J03552337$+$1133437	&L5	&14.05\,$\pm$\,0.02	&10.53\,$\pm$\,0.02	& 9.94\,$\pm$\,0.02	& 9.22\,$\pm$\,0.04	&--	                &y,y  & 50$-$500	& 70--120 & 15--40   &1,4,6\\
KPNO-Tau 4	                &M9.5	&15.00\,$\pm$\,0.04	&12.82\,$\pm$\,0.03	&12.36\,$\pm$\,0.03	&--	                &--	                &y,y  & $\sim1$         &         & 11--15   &2,8,9,10,11\\
CFHT-BD-Tau 4	                &M7	&12.17\,$\pm$\,0.02	& 9.83\,$\pm$\,0.02	& 8.99\,$\pm$\,0.02	& 7.02\,$\pm$\,0.02	&5.12\,$\pm$\,0.03	&y,y  & $\sim1$	        &         & $\sim67$ &3,4,6,8,12\\
2MASS J05012406$-$0010452	&L4	&14.98\,$\pm$\,0.03	&12.05\,$\pm$\,0.02	&11.52\,$\pm$\,0.02	&10.75\,$\pm$\,0.11	&--	                &y,y  & 50$-$500	&         & 13--45   &1,4,6\\
2MASS J15525906$+$2948485	&L0	&13.48\,$\pm$\,0.03	&11.55\,$\pm$\,0.02	&11.20\,$\pm$\,0.02	&10.62\,$\pm$\,0.05	&--	                &y,y  & $\ge$500	&         & $\ge65$  &1,4,6\\
2MASS J17260007$+$1538190	&L3	&15.67\,$\pm$\,0.07	&13.07\,$\pm$\,0.02	&12.67\,$\pm$\,0.03	&11.88\,$\pm$\,0.29	&--	                &y,y  & 10$-$300	&         & 11--45   &1,4,6\\
PSO J318.5338$-$22.8603	&L7	&16.71\,$\pm$\,0.20	&13.22\,$\pm$\,0.03	&12.46\,$\pm$\,0.03	&11.80\,$\pm$\,0.40	&--	                &n,y  & 	        & 8--20   & 6--15    &9\\
2MASS J21265040$-$8140293      &L3 	&15.54\,$\pm$\,0.06	&12.93\,$\pm$\,0.02	&12.47\,$\pm$\,0.02	&11.93\,$\pm$\,0.18	&--	                &y,n  &         	& 20--40  & $\sim$14      &1,13\\
2MASS J22081363$+$2921215	&L3	&15.80\,$\pm$\,0.09	&13.38\,$\pm$\,0.03	&12.91\,$\pm$\,0.03	&--	                &--	                &y,y  & 10$-$300	& 12--22  & 11--45   &1,4,6\\
2MASS J23225299$-$6151275	&L2	&15.54\,$\pm$\,0.06	&13.22\,$\pm$\,0.02	&12.81\,$\pm$\,0.03	&12.40\,$\pm$\,0.35	&--	                &y,n  & 	        & 20--40 & $\sim$13 &1,13\\
\hline\hline
\end{tabular} 

\begin{minipage}{175.5mm}
Notes: $^{\rm a}$ According to \cite{2014A&A...568A...6Z}, except for Taurus sources. $^{\rm b}$ Based on the objects' likely membership in star moving groups according to \citet{2013ApJ...777L..20L} and \citet{2014ApJ...783..121G}. References: (1) \cite{2009AJ....137.3345C}; (2) \cite{2004ApJ...600.1020M}; (3) \cite{2001ApJ...561L.195M}; (4) \cite{2013ApJ...772...79A}; (5) \cite{2009ApJ...697..824A};(6) \cite{2014A&A...568A...6Z}; (7) \cite{2012MNRAS.423.1775M}; (8) \cite{2013MNRAS.435.2650C}; (9) \cite{2013ApJ...777L..20L}; (10) \cite{2005ApJ...625..906M}; (11) \cite{2006ApJ...645.1498S}; (12) \cite{2009ApJ...697..373R}; (13) \cite{2014ApJ...783..121G}.
\end{minipage}
\end{table*}

\section{Target selection}\label{targets}

Our sample consists of 14 ultra-cool, relatively bright M7--L7 dwarfs ($12.1<\,J\,\le\,16.7$ mag). The common property of our targets is that they display spectroscopic features indicative of low- to intermediate-gravity atmospheres, thus suggesting young ages: weak \ion{Na}{ I} and \ion{K}{I} lines at optical and near-infrared wavelengths, strong absorption bands due to VO and H$_{2}$O, weak absorption by FeH, and a peaked shape of the $H$-band continuum \citep{2001MNRAS.326..695L,2009AJ....137.3345C, 2013ApJ...772...79A}. Some also show lithium absorption in their optical spectra (see below). Eleven targets were selected from the catalog of \citet{2009AJ....137.3345C}, representing $\sim48\,\%$ of the objects studied by these authors. The following three sources were added to the final target list: PSO\,J318.5338$-$22.8603 \citep{2013ApJ...777L..20L}, which is one of the reddest known field L dwarfs with a triangular-shaped $H$-band, and KPNO-Tau~4 \citep{2002ApJ...580..317B} and CFHT-BD-Tau~4 \citep{2001BCFHT..43...12D}, which are two confirmed members of the Taurus star-forming region.

\begin{figure}
\centering
\includegraphics[width=0.49\textwidth]{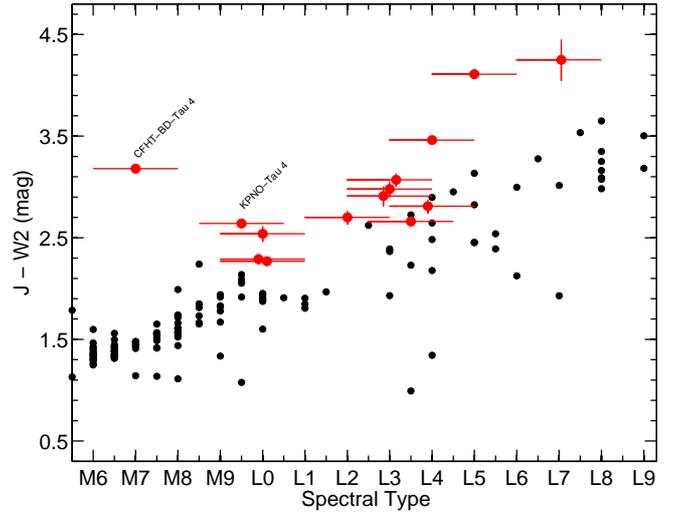}
\caption{$J-W2$ color as a function of spectral type. Our targets are plotted as red circles, vertical bars stand for errors in the color, while horizontal bars stand for the spectral type uncertainties. Field dwarfs taken from \citet{2012ApJS..201...19D} are shown with black circles. Taurus objects are labeled. Targets spectral types are slightly shifted for clarity.}
\label{Fig1}
\end{figure}

The complete names, 2MASS $J$-band and {\sl WISE} magnitudes \citep{2006AJ....131.1163S,2010AJ....140.1868W}, and spectral types of the targets are provided in Table~\ref{Table1} (abridged names will be used in what follows). We also indicate the availability of optical and near-infrared spectra in the literature. Regarding the {\sl WISE} photometry, we only provide magnitudes with signal-to-noise ratio (S/N) $\ge3$ as given by the  {\sl ALLWISE} catalog \citep{2013yCat.2328....0C}, except for PSO J318$-$22, whose values come from the discovery paper \citep{2013ApJ...777L..20L}. Figure~\ref{Fig1} shows the color $J-W2$ of the targets as a function of spectral type (the two Taurus objects are labeled). The sample systematically lies at very red colors defining the upper envelope of the trend delineated by field, high-gravity dwarfs. Among the target list, five sources (CFHT-BD-Tau~4 and KPNO-Tau~4; J0501$-$0010, L4; J0355$+$1133, L5; and PSO\,J318--22, L7) have clear mid-infrared flux excesses, deviating from the field sequence by more than 3-$\sigma$ the combined photometric errors and field-sequence dispersion.

%
%
%

The 14 targets are likely younger than $\approx$0.5-1 Gyr and less massive than $\approx$0.075 M$_\odot$. The youngest sources ($\approx$1 Myr) are the two confirmed Taurus members. For the remaining objects, we relied on the literature and on the presence of lithium in the atmospheres (i.e., the lithium test, \citealt{1992ApJ...389L..83R}) to estimate their ages and set a constraint on their masses. In the target sample, lithium absorption at 670.82 nm is present in J0032--4405 \citep{1999A&A...351L...5E},  J0045+1634, J0355+1133, J0501$-$0010, J1726$+$1538, J2126$-8140$, J2208$+$2921 \citep{2001ApJ...555..368S,2009AJ....137.3345C,2014A&A...568A...6Z}, and KPNO-Tau 4 \citep{2004ApJ...600.1020M,2005ApJ...626..498M}. Given their spectral types and likely young ages, all of these sources with atmospheric lithium have substellar masses likely ranging from 0.012 through 0.065 M$_\odot$. The upper limits on the lithium strength of J0033--1521, J0241--0326, J1552+2948, and J2322--6151 \citep{2009AJ....137.3345C,2014A&A...568A...6Z} indicate that these objects are likely older and have a mass greater than about 0.06 M$_\odot$. \citet{2013ApJ...772...79A} classified J0033--1521 as a normal field-gravity dwarf based on its near-infrared spectrum in agreement with the discussion by \citet{2014A&A...568A...6Z}. \citet{2014ApJ...783..121G} argued that J2322--6151 is a likely 0.012--0.013 M$_\odot$ member of the Tucana-Horologium moving group (20--40 Myr). We adopted these values since the spectrum of \citet{2009AJ....137.3345C} has poor quality at the lithium wavelengths, thus impeding the clear detection of the line. CFHT-BD-Tau~4 has no lithium detection reported in the literature to date; however, it shows several other indicators of extreme youth, like mid-infrared flux excesses and strong H$\alpha$ emission (suggesting mass accretion from a disk, e.g., \citealt{2003ApJ...592..282J}). PSO\,J318--22 has no optical spectrum. The age and mass intervals of our targets are summarized in Table~\ref{Table1} and respond to the measures determined by the references given in the last column. Using near-infrared, high spatial resolution images obtained with the {\sl Hubble Space Telescope}, \citet{2003AJ....125.3302G} and \citet{2006AJ....132..891R} did not resolve as binary the following targets: J0033$-$1521, J0045$+$1634, J0355$+$1133, J1552$+$2948, and J2208$+$2921. There are no other similar imaging studies to address the resolved binarity of the remaining targets in our sample.

Trigonometric parallaxes are available for nearly all of the targets in the sample \citep{2012ApJ...752...56F,2014A&A...568A...6Z}, except for J2126--8140, J2322--6151, and the Taurus objects. The latter are located at $\approx140$\,pc, which is the accepted distance to this expanded star-forming region \citep{1994AJ....108.1872K,1998MNRAS.301L..39W}, and they likely suffer from some extinction as reported by \citet{2007A&A...465..855G}. According to the measured parallactic distances, our field targets are typically located at $\lesssim$\,50\,pc. In the sample, J0045$+$1634 and J0241$-$0326 have $J$-band linear polarimetric data published in the literature \citep{2011ApJ...740....4Z,2013A&A...556A.125M}. The new measurements will allow us to monitor their polarimetric signal on different occasions.


\begin{table*}
\caption{Observing log.}
\label{Table2}
\renewcommand{\arraystretch}{1.0}
\setlength{\tabcolsep}{2.5pt}
\centering
\begin{tabular}{l c c c c c c c}
\hline\hline
Object& Instrument 	&Aperture$^{\rm a}$	& Obs. date 	&Filter 	&Exposure Time$^{\rm b}$ 	&FWHM		&Airmass\\
 	&	&(FWHM)			&(UT)		& 		& (s)			&(")	&	\\
\hline
J0032$-$4405&FORS2		&1$-$2		&2012 Aug 23	&$I$	&4$\,\times\,$1$\,\times\,$200								&1.3				&1.06$-$1.10\\
J0033$-$1521&FORS2		&1$-$2		&2012 Aug 23	&$I$	&2$\,\times\,$1$\,\times\,$250								&1.2				&1.18$-$1.33\\
J0045$+$1634&LIRIS	       &2$-$6		&2013 Oct 10	&$J$	&1$\,\times\,$9$\,\times\,$50, 1$\,\times\,$9$\,\times\,$50	&0.8				&1.10$-$1.06\\
                       &LIRIS	       &2$-$6		&2014 Oct 02	&$Y$&1$\,\times\,$9$\,\times\,$90, 1$\,\times\,$9$\,\times\,$90	&1.1				&1.09$-$1.05\\
                       &LIRIS	       &2$-$6		&2014 Oct 02	&$J$&1$\,\times\,$9$\,\times\,$30, 1$\,\times\,$9$\,\times\,$30	&1.1				&1.05$-$1.04\\
                       &ALFOSC	       &1$-$3		&2014 Oct 02	&$I$	&1$\,\times\,$1$\,\times\,$300								&1.0				&1.08$-$1.06\\
                       &ALFOSC	       &1$-$2		&2014 Oct 02	&$R$	&1$\,\times\,$1$\,\times\,$400								&1.0			&1.05$-$1.04\\
J0241$-$0326	&LIRIS	&1$-$2		&2011 Dec 31	&$J$	&1$\,\times\,$9$\,\times\,$120, 1$\,\times\,$9$\,\times\,$120	&1.2				&1.24$-$1.18\\
                       &FORS2 	&1$-$2		&2012 Aug 23	&$I$	&5$\,\times\,$1$\,\times\,$300								&0.9 			&2.30$-$1.30\\
			&LIRIS	&1.5$-$2.5	&2012 Oct 06	&$J$	&2$\,\times\,$9$\,\times\,$120, 2$\,\times\,$9$\,\times\,$120		&0.8				&1.21$-$1.19\\
			&LIRIS	&1.5$-$2.5	&2012 Oct 07	&$J$	&2$\,\times\,$9$\,\times\,$120, 2$\,\times\,$9$\,\times\,$120		&1.2				&1.25$-$1.18\\
			&LIRIS	&2$-$3    	&2013 Oct 12	&$J$	&2$\,\times\,$9$\,\times\,$60, 2$\,\times\,$9$\,\times\,$60		&0.9				&2.27$-$1.79\\
			&LIRIS	&1.5$-$2.5	&2013 Oct 13	&$Z$&3$\,\times\,$5$\,\times\,$60, 3$\,\times\,$5$\,\times\,$60		&0.8				&1.29$-$1.21\\
			&LIRIS	&2$-$3	&2013 Oct 13	&$J$	&7$\,\times\,$5$\,\times\,$60, 7$\,\times\,$5$\,\times\,$60		&0.7				&1.20$-$1.19\\
			&LIRIS	&2$-$4	&2013 Oct 13	&$H$	&2$\,\times\,$5$\,\times\,$40, 2$\,\times\,$5$\,\times\,$40		&0.7				&1.20$-$1.24\\
J0355$+$1133&LIRIS	       &2$-$6		&2013 Oct 10	&$J$	&1$\,\times\,$9$\,\times\,$50, 1$\,\times\,$9$\,\times\,$50	&0.8				&1.19$-$1.23\\
KPNO-Tau 4	&LIRIS 	&2$-$3		&2012 Oct 07	&$J$	&2$\,\times\,$9$\,\times\,$120, 2$\,\times\,$9$\,\times\,$120		&0.9				&1.06$-$1.00\\
			&LIRIS	&2$-$3		&2013 Jan 28	&$J$	&1$\,\times\,$9$\,\times\,$100, 1$\,\times\,$9$\,\times\,$100		&0.9				&1.28$-$1.42\\
			&LIRIS	&2$-$3		&2013 Oct 13	&$J$	&1$\,\times\,$9$\,\times\,$60, 1$\,\times\,$9$\,\times\,$60		&0.7				&1.09$-$1.13\\
CFHT-BD-Tau 4&LIRIS 	&4$-$6		&2013 Oct 12	&$Z$	&2$\,\times\,$9$\,\times\,$40, 2$\,\times\,$9$\,\times\,$40		&0.7				&1.07$-$1.15\\
                       &LIRIS 	&4$-$6		&2013 Oct 12	&$J$ 	&2$\,\times\,$9$\,\times\,$10, 2$\,\times\,$9$\,\times\,$10		&0.7				&1.03$-$1.06\\
J0501$-$0010	&LIRIS	&1$-$2		&2011 Dec 31	&$J$	&2$\,\times\,$9$\,\times\,$120, 2$\,\times\,$9$\,\times\,$120		&1.7				&1.16$-$1.78\\
J1552+2948	&LIRIS		&2$-$3		&2011 Dec 31	&$J$	&1$\,\times\,$9$\,\times\,$60, 1$\,\times\,$9$\,\times\,$60		&0.8				&1.97$-$1.68\\
			&LIRIS	&2$-$3		&2012 Jun 16	&$J$	&1$\,\times\,$9$\,\times\,$120, 2$\,\times\,$9$\,\times\,$120	&2.1				&1.01$-$1.16\\
			&LIRIS	&2$-$3    	&2013 Jan 28	&$J$	&1$\,\times\,$9$\,\times\,$80, 1$\,\times\,$9$\,\times\,$80		&0.8			&1.32$-$1.22\\
J1726+1538	&LIRIS	&1$-$2		&2012 Jun 15	&$J$	&2$\,\times\,$9$\,\times\,$120, 2$\,\times\,$9$\,\times\,$120	&1.1				&1.10$-$1.20\\
			&LIRIS	&1$-$2		&2013 Jan 28	&$J$	&1$\,\times\,$9$\,\times\,$120, 1$\,\times\,$9$\,\times\,$120		&1.1				&1.90$-$1.34\\
PSO J318$-$22	&LIRIS	&1$-$2		&2013 Oct 12	&$J$	&2$\,\times\,$9$\,\times\,$60, 2$\,\times\,$9$\,\times\,$60	&0.7					&1.62$-$1.65\\
                	&LIRIS	&2$-$3		&2013 Oct 11	&$K_{\rm s}$	&2$\,\times\,$9$\,\times\,$30, 2$\,\times\,$9$\,\times\,$30	&0.7					&1.74$-$1.67\\
J2126$-$8140	&FORS2	&1$-$2		&2012 Aug 23	&$I$	&5$\,\times\,$2$\,\times\,$200								&1.2				&2.10$-$1.90\\
J2208+2921	&LIRIS	&1$-$2		&2012 Jun 16	&$J$	&3$\,\times\,$9$\,\times\,$120, 2$\,\times\,$9$\,\times\,$120	&1.7				&1.23$-$1.02\\
J2322$-$6151	&FORS2	&2$-$3 	&2012 Aug 23	&$I$	&2$\,\times\,$3$\,\times\,$200, 4$\,\times\,$1$\,\times\,$200	&1.4				&1.54$-$1.29\\
\hline
Feige 110$^{\rm c}$        &LIRIS&2$-$4	&2013 Oct 11	&$Z$		&1$\times$5$\times$20, 1$\times$5$\times$20	&0.7 		&1.22--1.22\\
              				&LIRIS&2$-$4	&2013 Oct 11	&$J$		&1$\times$9$\times$10, 1$\times$9$\times$10	&0.7			&1.25--1.25\\
              				&LIRIS&2$-$4	&2013 Oct 11	&$H$		&1$\times$5$\times$12, 1$\times$5$\times$12	&0.7			&1.24--1.24\\
              				&LIRIS&3$-$4	&2013 Oct 13	&$Z$		&1$\times$5$\times$30, 1$\times$5$\times$30	&0.7			&1.29--1.30\\
              				&LIRIS&3$-$4	&2013 Oct 13	&$J$		       &2$\times$5$\times$10, 2$\times$5$\times$10	&0.7 		&1.27--1.29\\
              		        	&LIRIS&3$-$4	&2013 Oct 13	&$H$		&1$\times$5$\times$7, 1$\times$5$\times$7	&0.7			&1.25--1.25\\
              				&LIRIS&3$-$4	&2013 Oct 13	&$K_{\rm s}$		&2$\times$5$\times$7, 2$\times$5$\times$7	&0.7 		&1.26--1.27\\
BD$+28$\,4211$^{\rm c}$&LIRIS		&1$-$6		&2014 Oct 02	&$Y$&1$\,\times\,$9$\,\times\,$7, 1$\,\times\,$9$\,\times\,$7		&0.8	&1.15$-$1.14\\
                                          &LIRIS		&1$-$6		&2014 Oct 02	&$J$&1$\,\times\,$9$\,\times\,$7, 1$\,\times\,$9$\,\times\,$7		        &0.8		&1.13$-$1.12\\
WD 0310$-$688$^{\rm c}$&FORS2	&2$-$6		&2012 Aug 23	&$I$	&1$\,\times\,$1$\,\times\,$0.25		&0.4		&1.38$-$1.39\\
WD 0501$+$527$^{\rm c}$&ALFOSC	&2$-$6		&2014 Oct 02	&$R$	&4$\,\times\,$2$\,\times\,$3		&0.8		&1.09$-$1.09\\
                                         &ALFOSC	&2$-$6		&2014 Oct 02	&$I$	       &4$\,\times\,$2$\,\times\,$5		&0.8 	&1.09$-$1.09\\
Hiltner 652$^{\rm d}$&FORS2		&2$-$6		&2012 Aug 23	&$I$	&2$\,\times\,$2$\,\times\,$0.25		&1.2		&1.02$-$1.01\\
HD 38563c$^{\rm d}$&LIRIS		&3$-$6		&2013 Oct 10	&$J$	&2$\,\times\,$9$\,\times\,$3, 2$\,\times\,$9$\,\times\,$3		&0.8		&1.17$-$1.18\\
                                  &LIRIS		&3$-$6		&2013 Oct 10	&$H$	&2$\,\times\,$9$\,\times\,$2, 2$\,\times\,$9$\,\times\,$2		&0.8		&1.18$-$1.19\\
                                  &LIRIS		&3$-$4		&2013 Oct 12	&$Z$	&1$\,\times\,$5$\,\times\,$5, 1$\,\times\,$5$\,\times\,$5		&0.7		&1.18$-$1.18\\
                                  &LIRIS		&2$-$6		&2013 Oct 12	&$J$	       &1$\,\times\,$5$\,\times\,$2, 1$\,\times\,$5$\,\times\,$2		        &0.7		&1.19$-$1.19\\
                                  &LIRIS		&2$-$6		&2013 Oct 12	&$H$	&1$\,\times\,$5$\,\times\,$1, 1$\,\times\,$5$\,\times\,$1		&0.7		&1.20$-$1.20\\
HD 283855$^{\rm d}$&LIRIS		&1$-$6		&2014 Oct 02	&$Y$&1$\,\times\,$9$\,\times\,$2, 1$\,\times\,$9$\,\times\,$2		&0.8			&1.01$-$1.01\\
                                &LIRIS		&1$-$6		&2014 Oct 02	&$J$&1$\,\times\,$9$\,\times\,$1, 1$\,\times\,$9$\,\times\,$1		        &0.8		&1.01$-$1.01\\
BD$+$64\,106$^{\rm d}$&ALFOSC		&2$-$6		&2014 Oct 02	&$R$	&4$\,\times\,$2$\,\times\,$1		&1.2		&1.62$-$1.63\\
                                       &ALFOSC		&2$-$6		&2014 Oct 02	&$I$ 	&4$\,\times\,$2$\,\times\,$1		&1.2		&1.63$-$1.65\\
\hline\hline
\end{tabular}
\begin{minipage}{175.5mm}
Notes: $^{\rm a}$ Range of circular photometric apertures (in units of FWHM) used to determine the normalized Stokes parameters. $^{\rm b}$ Total integration times per instrument and observing run as explained in the text. $^{\rm c}$ Zero-polarized standard star. $^{\rm d}$ Polarized standard star \citep{1992ApJ...386..562W}.
\end{minipage}
\end{table*}

\section{Observations and data reduction \label{observations}}

The log of optical and near-infrared linear polarimetry observations is provided in Table\,\ref{Table2}, including the total on-source integration times, Universal Time (UT) observing dates, filters, mean air masses, and raw seeing as measured from the averaged full-width-at-half maximum (FWHM) over the reduced images. Further details on data acquisition and reduction are given in the following subsections.

\subsection{Optical linear polarimetry}

\subsubsection{FORS2}

The FOcal Reducer and low dispersion Spectrograph (FORS2) sample contains five objects (J0032$-$4405, J0033$-$1521, J0241$-$0326, J2126$-$8140, and J2322$-$6151). We conducted linear polarimetric imaging photometry using the Bessel $I$-band filter and FORS2 \citep{1998Msngr..94....1A} on the Antu unit (UT1) of the Very Large Telescope (VLT) of the European Southern Observatory (ESO) in Cerro Paranal, Chile. FORS2 is by default equipped with a detector system that is optimized for the red with a very low level of fringes thanks to a mosaic of two 2048$\,\times\,$4096 MIT CCDs (with 15 $\mu$m pixels). The plate scale is 0$\farcs$252 pix  for the standard readout mode (2$\times$2 binning). The polarization optics consists of a Wollaston prism as beam splitting analyzer and a half-wave superachromatic retarder plate to measure linear polarization. When using the imaging polarimetry mode, a mask with alternating transparent and opaque parallel strips avoids the overlapping of the two beams from the Wollaston, yielding half of the field of view, which is duplicated as ordinary and extraordinary rays in strips of size 3$\farcm4\,\times\,11\arcsec$. The central wavelength and the passband of the Bessel $I$ filter are 0.768 and 0.138 $\mu$m. Observations were carried out on 2012 August 23. Data were acquired at $\ge90$ deg from the wind direction, which was pointing from the North at a velocity of $\ge$12 m\,s$^{-1}$. Weather conditions were clear and the seeing was around 1$\farcs$1 during the whole night. FORS2 polarimetric images were acquired using different positions of the phase retarder plate (0, 22.5, 45, and 67.5 deg). One polarimetric cycle includes the four angles. One to three images were taken per retarder plate angle (Table\,\ref{Table2}). To avoid as many systematics as possible, all targets were acquired on the same spot of the FORS2 detector 1 (no dithering). Individual exposure times ranged from 200 to 300 s depending on the target brightness. In Table~\ref{Table2}, total integration times are given as follows: number of polarimetric cycles $\times$ number of exposures per retarder plate $\times$ individual exposure time.

\subsubsection{ALFOSC}

We collected $R$- and $I$-band linear polarimetry images of J0045$+$1634 using the Andaluc\'ia Faint Object Spectrograph and Camera (ALFOSC\footnote[1]{ ALFOSC is provided by the Instituto de Astrof\'isica de Andaluc\'ia (IAA) under a joint agreement with the University of Copenhagen and the Nordic Optical Telescope Scientific Association (NOTSA).}) mounted on the 2.56-m Nordic Optical Telescope (NOT) of the Roque de los Muchachos Observatory (La Palma, Spain) on 2014 October 2. ALFOSC has a 2048$\times$2048 E2V detector with a pixel size of 0\farcs19. The central wavelength and the passband of the filters are 0.641/0.148  $\mu$m ($R$) and 0.797/0.157 $\mu$m ($I$). ALFOSC linear polarimetry mode consists of a half-wave plate and a calcite block providing simultaneous images of the ordinary and the extraordinary beams separated by 15\arcsec. The total unvignetted field of view is 140\arcsec~in diameter. Sky conditions were clear and the seeing was nearly constant around 1\arcsec~during the observations. The ALFOSC and the near-infrared data of  J0045$+$1634 were acquired simultaneously (see Section~\ref{nirp}). Similarly to FORS2, we employed four angles of the retarder plate (0, 22.5, 45, and 67.5 deg) to complete one polarimetric cycle of observations. ALFOSC individual exposure times were 400 s ($R$) and 300 s ($I$). 

In Table~\ref{Table2}, total times are given as follows: number of polarimetric cycles $\times$ number of exposures per retarder plate $\times$ individual exposure time. 

FORS2 and ALFOSC raw data were bias and flat-field corrected using packages within the Image Reduction and Analysis Facility software (IRAF\footnote[2]{IRAF is distributed by the National Optical Astronomy Observatories, which are operated by the Association of Universities for Research in Astronomy, Inc., under cooperative agreement with the National Science Foundation.}). We used images taken without the polarimetric optics during the sunset to construct proper flat-field images per observing filter.

\subsection{Infrared linear polarimetry (LIRIS) \label{nirp}}

Ten targets (J0045$+$1634, J0241$-$0326, J0355$+$1133, KPNO-Tau 4, CFHT-BD-Tau 4,  J0501$-$0010, J1552$+$2948, J1726$+$1538, PSO J318--22, and J2208$+$2921) were observed in the near-infrared using the $J$-band filter and the Long-slit Intermediate Resolution Infrared Spectrograph (LIRIS; \citealt{2004SPIE.5492.1094M}) attached to the Cassegrain focus of the 4.2-m William Herschel Telescope (WHT) on the Roque de los Muchachos Observatory. Some of them were also observed in the $Z$, $Y$, $H$, and $K_{\rm s}$ filters. LIRIS has a 1024\,$\times$\,1024 pixel Hawaii detector covering the spectral range 0.8--2.5 $\mu$m. The pixel projection on the sky is 0\farcs25 yielding a field of view of $4\farcm27\times4\farcm27$. In its polarimetric imaging mode, LIRIS uses a Wedged double Wollaston device \citep{1997A&AS..123..589O}, consisting in a combination of two Wollaston prisms that deliver four simultaneous images of the polarized flux at vector angles 0 and 90 deg, 45 and 135 deg. An aperture mask 4\arcmin$\,\times\,$1\arcmin~in size is in the light path to prevent overlapping effects between the different polarization vector images. The central wavelengths and widths of the LIRIS filters are $1.02/0.04$ $\mu$m ($Y$-band), 1.03$/$0.04 $\mu$m ($Z$-band), 1.25$/$0.16 $\mu$m ($J$-band), 1.63$/$0.15 $\mu$m ($H$-band), and 2.15$/$0.16 $\mu$m ($K_{\rm s}$-band). We obtained linear polarimetric images following a dither pattern of 5 or 9 points to properly remove the sky background contribution. Typical dither offsets were $\sim$20\arcsec~and $\sim$10\arcsec~along the horizontal and vertical axis. We systematically located our targets on the same spot of the detector, which is close to the center of the LIRIS field of view and optical axis.  Observations were carried out during six observing campaigns between 2011 December and 2014 October. Weather conditions were clear and the raw seeing was typically $0\farcs6$--$1\farcs3$, except for 2012 June when we had a seeing of $\sim2\arcsec$. 

A detailed description of  the LIRIS polarimetric frames obtained through the two Wollaston prisms is given in \citet{2011AJ....142...33A}: Per frame, there are four images of the main source corresponding to vector angles of 0$^{\circ}_{TR}$, 90$^{\circ}_{TR}$, 135$^{\circ}_{TR}$, and 45$^{\circ}_{TR}$ from top to bottom, where the subindex TR stands for the telescope rotator angle. The early observations of 2011 December, 2012 June, and 2012 October were acquired at two different positions of the WHT rotator: TR = 0$^{\circ}$ and 90$^{\circ}$. The benefits of this observing strategy are twofold: The flat-fielding effects of the detector are minimized, and S/N of the polarimetric measurements is improved. The overheads introduced by the rotation and de-rotation of the telescope rotator were typically about 5 min per target. For the observing runs in 2013 and 2014, we used two retarder plates, which is an implementation added to one of the filter wheels of LIRIS during late 2012. These two retarder plates minimize the overhead times (by not having to move the telescope rotator) and provide polarimetric images with exchanged orthogonal vector angles. Total integration times are listed per telescope rotator angle or retarder plate in Table~\ref{Table2} as follows: number of polarimetric cycles $\times$ dithering pattern $\times$ individual exposure time. 

Raw LIRIS data were reduced as in \citet{2013A&A...556A.125M}. Each polarimetric vector was processed separately. Data were sky subtracted, flat-fielded, aligned, and stacked together to produce deep images. The final S/N of our targets in the reduced FORS2, ALFOSC, and LIRIS images is $\ge$ 100.

\subsection{Linear polarization standard stars}

To control the instruments efficiency and the linear polarization introduced by the various telescopes and instruments, one polarized and one unpolarized standard stars were observed with the same instrumental configurations and on the same dates as the science objects. The log of the standard stars observations is given in Table~\ref{Table2} (bottom). These sources were selected from the catalogs of standard stars of FORS2 and ALFOSC (optical) and from  \citet{1992AJ....104.1563S} and \citet[][near infrared]{1992ApJ...386..562W}. In the case of FORS2, standards are too bright for an 8-m class telescope, so images were acquired with a bad active optics performance and using very short integration times to avoid saturation and the non-linear regime of the detector. All data of standard stars were reduced and analyzed in the same manner as the science targets. ALFOSC standard stars were observed using 16 angles of the retarder plate (this is 12 retarder plate angles in addition to the four angles employed for the science targets). This yields four independent measures of the linear polarization by considering groups of 4 consecutive angles, or one single measurement (with improved S/N) by combining all 16 angles of the retarder plate.

%

\section{Polarimetric analysis}\label{polarimetry}
We used the flux-ratio method to compute the normalized Stokes parameters $q$ and $u$ at optical and near infrared wavelengths. The degree of linear polarization, $P$, and the vibration angle of the polarization, $\Theta$, were obtained for all filters using measured fluxes and equations 5 and 6 given in \citet{2011ApJ...740....4Z}. We followed the procedure fully described in \citet{2013A&A...556A.125M}, which is summarized as follows: Fluxes of all polarimetric quantities involved in these equations were measured using the IRAF PHOT package and defining circular photometric apertures of different sizes (from 0.5\,$\times$ to 6\,$\times$\,FWHM with steps of 0.1\,$\times$\,FWHM) and 18 sky rings or annulus of inner radius of 3.5\,$\times$ through 6\,$\times$\,FWHM (steps of 0.5\,$\times$\,FWHM) and widths of 1\,$\times$, 1.5\,$\times$, and 2\,$\times$\,FWHM. We computed the Stokes parameters for each aperture and sky ring and plotted them as a function of aperture, then we chose those apertures where both $q$ and $u$ parameters remained flat (typically 2$\times$\,through 4$\times$\,FWHM). To illustrate this technique, we provide $q$ and $u$ determinations as a function of photometric aperture in Figure \ref{figap} in the Appendix \ref{ap1}. We finally picked the average $q$ and $u$ values of the selected apertures range. Their associated uncertainties were determined as the standard deviation of all measurements within the selected apertures. To test the validity of this method of computing the polarimetric uncertainties, we derived the $q$ and $u$ values for each of the 9 dither images of KPNO-Tau 4, which resulted in 9 independent $q$ and $u$ measurements. The standard deviation of these $q$ and $u$ values divided by $\sqrt{9}$ coincides with the quoted uncertainties determined for the combined image using the approach of \citet{2013A&A...556A.125M}. Tables~\ref{Table3} and \ref{Table4} provide the final normalized Stokes parameters and their error bars for both science targets and standard stars. 

From the observations of the non-polarized standard star using FORS2, we determined that the Bessel $I$-band instrumental linear polarization is inconsequential within $\pm\,0.04\%$, which agrees with the value of $\pm\,0.03\%$ reported by \citet{2007ASPC..364..503F}. From the  polarized standard star, we measured an offset in the linear polarization vibration angle of $\Theta_{\rm o}=-2\fdg69\,\pm\,0\fdg23$ for FORS2. This result agrees with the angle $\Theta_{\rm o,tab}=-2\fdg89\,\pm\,0\fdg10$ given in the instrument users manual\footnote[3]{http://www.eso.org/sci/facilities/paranal/instruments/fors/doc/}. 

Similarly, we determined the upper limit on the instrumental linear polarization and the correction for the polarization vibration angle of the ALFOSC instrument. In analogy to the science targets, the ALFOSC four retarder plate angles of the non-polarized star yielded no significant instrumental polarization within $\pm\,0.22\%$ for both the $R$- and $I$-bands, which agrees with the ALFOSC manual. A more precise determination was obtained from the combination of the 16 angles of the retarder plate, from which we constrained the instrumental linear polarization to be less than $\pm0.07\%$ for both filters of interest. The ALFOSC zero-points of the linear polarization vibration angle were measured at $\Theta_{\rm o}=\,2\fdg4\pm\,1\fdg8$ and $\Theta_{\rm o}=\,2\fdg2\pm\,1\fdg8$ for the $R$- and $I$-bands, respectively. 

As for LIRIS, we found negligible instrumental linear polarization within $\pm\,0.10\,\%$ for the $Y$-, $Z$-, $J$-, $H$-, and $K_{\rm s}$-filters, and vibration angle zero points depending on wavelength as follows: $\Theta_{\rm o} =3\fdg3\pm1\fdg6$ ($Y$), $4\fdg4\pm1\fdg2$ ($Z$), $4\fdg1\pm1\fdg3$ ($J$), and $5\fdg3\pm1\fdg5$ ($H$). These values coincide within 1-$\sigma$ the quoted uncertainties with those given by \citet{2013A&A...556A.125M} for the filters in common.

Tables~\ref{Table3} and~\ref{Table4} provide the linear polarization degrees, $P = (q^2 + u^2)^{1/2}$, and polarization vibration angles, $\Theta = 0.5\,{\rm atan}\,(u/q)$, determined for the science targets and standard stars. Error bars of $P$, $\sigma_P$, were computed as the quadratic sum of the $q$ and $u$ uncertainties plus the error in the determination of the instrumental polarization. The errors associated with $\Theta$ were computed as the sum of the error coming from the corresponding $\Theta_{\rm o}$ and the value derived from the expression $28.65\sigma_{P}/P$ (deg), which results from the error propagation theory and is valid when $P/\sigma_{P}\ge3$ \citep{1974apoi.book..361S,1974ApJ...194..249W}. Because the linear polarization is always a positive quantity, small values of $P$ and values of $P$ affected by poor S/N data are statistically biased toward an overestimation of the true polarization \citep[see][]{1985A&A...142..100S}. We applied the equation 
$p^{*}\,=\,(P^{2}-\sigma_{P}^{2})^{1/2}$
given by \citet{1974ApJ...194..249W} to derive the unbiased linear polarization degree, $p^{*}$ (also listed in Tables~\ref{Table3} and~\ref{Table4}), by taking into account the measured $P$ and its associated uncertainty. At high values of polarization and small uncertainties, differences between $P$ and $p^{*}$ are imperceptible. For those measurements where $\sigma_P \ge P$, we computed the 1-$\sigma$ upper limit for $p^{*}$ following \citet[][see their section 3]{1985A&A...142..100S} to indicate that the likely linear polarization value lies somewhere between 0.0 $\%$ and the computed upper limit with a confidence at the 67\%~level.

We adopted the 3-$\sigma$ criterion  (i.e.,  $P/\sigma_P\,\ge\,3$) for claiming detections of linear polarization with a significant confidence of 99\%. For these particular cases, we computed the polarization vibration angles and their associated uncertainties provided in Tables~\ref{Table3} and ~\ref{Table4}. The polarimetric data complying with the 3-$\sigma$ criterion are conveniently highlighted in all Figures throughout the paper. The polarization vibration angles were also determined for less significance polarimetric measurements ($1 < P/\sigma_p < 3$) and are only shown in  Figures of Sections~\ref{pt} and~\ref{wav}.

\begin{table*}
\caption{Linear polarimetry photometry of science targets.}\label{Table3}
\renewcommand{\arraystretch}{1.0}
\setlength{\tabcolsep}{4pt}
\centering
\begin{tabular}{l c c r r r c r} 
\hline\hline             
Object	 	&Filter&Obs$.$ date			&\multicolumn{1}{c}{$q$}	&\multicolumn{1}{c}{$u$}	&\multicolumn{1}{c}{$P$}	&\multicolumn{1}{c}{$p^{*}$}	&\multicolumn{1}{c}{$\Theta$}\\
  			& &(JD$-$2450000.5) 	&\multicolumn{1}{c}{($\%$)}	&\multicolumn{1}{c}{($\%$)}	&\multicolumn{1}{c}{($\%$)}	&\multicolumn{1}{c}{($\%$)}	&\multicolumn{1}{c}{(deg)} \\
\hline
\hline
\multicolumn{8}{c}{LIRIS data}\\
\hline
J0045$+$1634     &$J$	&6575.9719	&$-0.23\,\pm\,0.04$	&$0.07\,\pm\,0.10$	&$0.24\,\pm\,0.15$	&$0.19\,\pm\,0.15$	&    \\
                            &$R^{\rm a}$	&6933.0173	&$0.16\,\pm\,0.32$	&$-0.08\,\pm\,0.18$	       &$0.17\,\pm\,0.43$	&$\,\le\,$0.30	&    \\
                            &$I^{\rm a}$	&6933.0006	&$0.04\,\pm\,0.19$	&$0.13\,\pm\,0.22$	       &$0.14\,\pm\,0.37$	&$\,\le\,$0.24	&    \\
                            &$Y$	&6932.9991	&$0.31\,\pm\,0.08$	&$0.09\,\pm\,0.60$	       &$0.32\,\pm\,0.14$	&$0.29\,\pm\,0.14$	&    \\
                            &$J$	&6933.0116	&$0.32\,\pm\,0.05$	&$0.21\,\pm\,0.04$	       &$0.38\,\pm\,0.12$	&$0.36\,\pm\,0.12$	&16.6$\,\pm\,$9.5   \\
J0241$-$0326		&$J$	&5926.8583	&$-0.60\,\pm\,0.09$	&$0.39\,\pm\,0.17$	&$0.72\,\pm\,0.22$	&$0.69\,\pm\,0.22$	&73.5$\,\pm\,$8.7	\\
				&$J$	&6207.0830	&$1.39\,\pm\,0.15$	&$-0.03\,\pm\,0.10$	&$1.39\,\pm\,0.21$	&$1.38\,\pm\,0.21$	&179.4$\,\pm\,$4.3\\
				&$J$	&6208.0938	&$1.16\,\pm\,0.15$	&$-0.32\,\pm\,0.17$	&$1.20\,\pm\,0.25$	&$1.18\,\pm\,0.25$	&172.3$\,\pm\,$5.9\\
                           	&$J$	&6577.9558	&$-0.05\,\pm\,0.13$	&$-0.14\,\pm\,0.23$	&$0.15\,\pm\,0.29$	& $\,\le\,$0.26	&    \\
                           	&$Z$&6579.0584	&$-1.00\,\pm\,0.77$	&$3.83\,\pm\,0.46$	&$3.96\,\pm\,0.91$	&$3.86\,\pm\,0.91$	&52.3$\,\pm\,$6.6    \\
                               &$J$	&6579.0721   &$-0.06\,\pm\,0.16$ 	&$0.12\,\pm\,0.12$	       &$0.14\,\pm\,0.22$	&$\,\le\,$0.22 	& \\
      			       &$J$	&6579.0808   &$-0.55\,\pm\,0.12$	        &$0.98\,\pm\,0.13$	       &$1.12\,\pm\,0.20$	&$1.11\,\pm\,0.20$	&59.7$\,\pm\,$5.2 \\
        		       &$J$	&6579.0889   &$-0.02\,\pm\,0.22$	        &$0.39\,\pm\,0.10$	       &$0.39\,\pm\,0.27$	&$0.28\,\pm\,0.27$	& \\
    			      &$J$	&6579.0971   &$-0.47\,\pm\,0.13$ 	&$-0.56\,\pm\,0.20$	       &$0.73\,\pm\,0.25$	&$0.68\,\pm\,0.25$	& \\
    			      &$J$	 &6579.1056   &$0.27\,\pm\,0.25	$        &$-0.81\,\pm\,0.11$ 	&$0.85\,\pm\,0.29$	&$0.80\,\pm\,0.29$	& \\
     			      &$J$	 &6579.1139   &$-0.01\,\pm\,0.15$	         &$0.16\,\pm\,0.27$	       &$0.16\,\pm\,0.33$	&$\,\le\,$0.30	& \\
   			      &$J$	&6579.1221   &$0.65\,\pm\,0.11$        &$-1.30\,\pm\,0.28$	      &$1.45\,\pm\,0.32$	              &$1.42\,\pm\,0.32$	&148.3$\,\pm\,$6.3 \\
                           	&$H$	&6579.1322  &$0.04\,\pm\,0.13$	&$0.12\,\pm\,0.23$	&$0.13\,\pm\,0.28$	&$\,\le\,$0.23	&   \\
J0355$+$1133     	&$J$	&6576.2404	&$0.06\,\pm\,0.10$	&$-0.41\,\pm\,0.08$	&$0.42\,\pm\,0.16$	&$0.38\,\pm\,0.16$	&    \\
KPNO-Tau 4     	&$J$	&6208.1250	&$0.86\,\pm\,0.08$	&$0.31\,\pm\,0.11$	&$0.91\,\pm\,0.17$	&$0.90\,\pm\,0.17$	&9.9$\,\pm\,$5.3\\
				&$J$	&6321.0116	&$-0.16\,\pm\,0.16$	&$0.43\,\pm\,0.13$	&$0.46\,\pm\,0.23$	&$0.40\,\pm\,0.23$	&	\\
				&$J$	&6579.2441	&$-0.31\,\pm\,0.17$	&$-0.73\,\pm\,0.08$	&$0.79\,\pm\,0.22$	&$0.76\,\pm\,0.22$	&123.5$\,\pm\,$7.7\\
CFHT-BD-Tau 4     	&$J$	&6578.2314	&$-0.61\,\pm\,0.04$	&$0.24\,\pm\,0.06$	&$0.65\,\pm\,0.12$	&$0.64\,\pm\,0.12$	& 79.3$\,\pm\,$5.4\\
                          	&$Z$	&6578.2516	&$-0.52\,\pm\,0.06$	&$0.22\,\pm\,0.08$	&$0.56\,\pm\,0.17$	&$0.53\,\pm\,0.17$	& 78.5$\,\pm\,$8.7\\
J0501--0010     	&$J$	&5927.0120	&$0.19\,\pm\,0.09$	&$0.23\,\pm\,0.07$	&$0.30\,\pm\,0.16$	&$0.25\,\pm\,0.16$	&	\\
J1552+2948	       &$J$	&5927.2534	&$0.20\,\pm\,0.04$	&$0.11\,\pm\,0.07$	&$0.23\,\pm\,0.13$	&$0.19\,\pm\,0.13$	&	\\
	                      &$J$	&6094.9531	&$-0.06\,\pm\,0.05$	&$-0.05\,\pm\,0.04$	&$0.08\,\pm\,0.12$	&$\,\le\,$0.12&	\\
				&$J$	&6321.2332	&$0.23\,\pm\,0.08$	&$0.42\,\pm\,0.08$	&$0.48\,\pm\,0.15$	&$0.45\,\pm\,0.15$	&30.6$\,\pm\,$9.0\\
J1726+1538	      &$J$	&6094.1205	&$-0.28\,\pm\,0.09$	&$0.34\,\pm\,0.08$	&$0.44\,\pm\,0.15$	&$0.41\,\pm\,0.15$	&	\\
				&$J$	&6321.2732	&$0.18\,\pm\,0.09$	&$0.69\,\pm\,0.15$	&$0.72\,\pm\,0.20$	&$0.69\,\pm\,0.20$	&37.7$\,\pm\,$8.1	\\
PSO J318$-$22     &$J$&6577.9001	&$-0.06\,\pm\,0.15$	&$0.42\,\pm\,0.42$	&$0.43\,\pm\,0.45$	&$\,\le\,$0.72&    \\
                        	&$K_{\rm s}$	&6576.8264&$-0.18\,\pm\,0.10$	&$0.35\,\pm\,0.22$	&$0.39\,\pm\,0.30$	&$0.25\,\pm\,0.30$	&    \\
J2208+2921	      &$J$	&6095.1193	&$-0.23\,\pm\,0.13$	&$0.06\,\pm\,0.14$	&$0.24\,\pm\,0.21$	&$0.11\,\pm\,0.21$	&	\\
\hline
\multicolumn{8}{c}{FORS2 data}\\
\hline
J0032--4405			&$I$	&6163.2989	&$-0.05\,\pm\,0.09$	&$-0.14\,\pm\,0.10$	&$0.15\,\pm\,0.14$	&$0.05\,\pm\,0.14$	&	\\
					&$I$	&6163.3109	&$-0.14\,\pm\,0.07$	&$-0.08\,\pm\,0.09$	&$0.16\,\pm\,0.12$	&$0.11\,\pm\,0.12$	&	\\
					&$I$	&6163.3224	&$-0.24\,\pm\,0.11$	&$-0.03\,\pm\,0.09$	&$0.25\,\pm\,0.14$	&$0.20\,\pm\,0.14$	&	\\
					&$I$	&6163.3458	&$-0.04\,\pm\,0.05$	&$-0.01\,\pm\,0.11$	&$0.04\,\pm\,0.13$	&$\,\le\,$0.10	&	\\
J0033-1521 			&$I$	&6163.3930	&$-0.10\,\pm\,0.03$	&$-0.04\,\pm\,0.07$	&$0.11\,\pm\,0.09$	&$0.06\,\pm\,0.09$	&	\\
		 			&$I$	&6163.4068	&$0.03\,\pm\,0.05$	&$0.00\,\pm\,0.08$	&$0.03\,\pm\,0.09$	&$\,\le\,$0.10	&	\\
J0241-0326			&$I$	&6163.2139	&$0.84\,\pm\,0.21$	&$0.26\,\pm\,0.28$	&$0.88\,\pm\,0.35$	&$0.80\,\pm\,0.35$	&	\\
					&$I$	&6163.2299	&$0.56\,\pm\,0.20$	&$0.29\,\pm\,0.27$	&$0.63\,\pm\,0.34$	&$0.53\,\pm\,0.28$	&	\\
					&$I$	&6163.2462	&$0.47\,\pm\,0.13$	&$0.12\,\pm\,0.29$	&$0.48\,\pm\,0.32$	&$0.36\,\pm\,0.32$	&	\\
					&$I$	&6163.2624	&$-0.08\,\pm\,0.22$	&$0.11\,\pm\,0.19$	&$0.14\,\pm\,0.30$	&$\,\le\,$0.24	&	\\
					&$I$	&6163.2790	&$0.87\,\pm\,0.16$	&$1.15\,\pm\,0.25$	&$1.44\,\pm\,0.30$	&$1.41\,\pm\,0.30$	&26.4$\,\pm\,$6.0	\\
J2126-8140 			&$I$	&6163.0059	&$0.62\,\pm\,0.13$	&$-0.49\,\pm\,0.22$	&$0.79\,\pm\,0.25$	&$0.75\,\pm\,0.25$	&160.8$\,\pm\,$9.1\\
   					&$I$	&6163.0277	&$0.12\,\pm\,0.19$	&$-0.15\,\pm\,0.21$	&$0.19\,\pm\,0.28$	&$\,\le\,$0.30	& 	\\
   					&$I$	&6163.0502	&$-0.02\,\pm\,0.10$	&$-0.04\,\pm\,0.17$	&$0.04\,\pm\,0.19$	&$\,\le\,$0.10	& 	\\
   					&$I$	&6163.0721	&$0.40\,\pm\,0.10$	&$0.06\,\pm\,0.12$	&$0.40\,\pm\,0.15$	&$0.37\,\pm\,0.15$	& 	\\
   					&$I$	&6163.0941	&$0.82\,\pm\,0.13$	&$0.06\,\pm\,0.10$	&$0.82\,\pm\,0.16$	&$0.80\,\pm\,0.16$	&2.1$\,\pm\,$5.6 	\\
J2322-6151$^{\rm b}$ 	&$I$	&6163.1260	&$0.23\,\pm\,0.07$	&$0.00\,\pm\,0.14$	&$0.23\,\pm\,0.16$	&$0.17\,\pm\,0.16$	&	\\
		 			&$I$	&6163.1586	&$0.22\,\pm\,0.10$	&$0.02\,\pm\,0.15$	&$0.22\,\pm\,0.18$	&$0.13\,\pm\,0.18$	&	\\
		 			&$I$	&6163.1806	&$0.10\,\pm\,0.13$	&$0.01\,\pm\,0.14$	&$0.10\,\pm\,0.16$	&$\,\le\,$0.17	&	\\
		 			&$I$	&6163.1926	&$0.16\,\pm\,0.12$	&$0.03\,\pm\,0.14$	&$0.16\,\pm\,0.18$	&$\,\le\,$0.27	&	\\
		 			&$I$	&6163.3758	&$0.11\,\pm\,0.11$	&$0.05\,\pm\,0.13$	&$0.10\,\pm\,0.17$	&$\,\le\,$0.17&	\\
\hline
\hline
\end{tabular}
\begin{minipage}{175.5mm}
Notes: $^{\rm a}$ ALFOSC data. $^{\rm b}$ Six polarimetric cycles were observed. The one dramatically affected by a cosmic ray is not shown.
\end{minipage}
\end{table*}


\begin{table*}
\caption{Linear polarimetry photometry of standard stars.}
\label{Table4}
\renewcommand{\arraystretch}{1.0}
\setlength{\tabcolsep}{4pt}
\centering
\begin{tabular}{l c c r r r c c}
\hline\hline
\multicolumn{8}{c}{{Unpolarized standard stars}}\\
\hline
Object			&Obs$.$ date		&Filter	 &\multicolumn{1}{c}{$q$}				&\multicolumn{1}{c}{$u$}		&\multicolumn{1}{c}{$P$}	&\multicolumn{1}{c}{$p^{*}$}	&\multicolumn{1}{c}{$\Theta$}\\
				&(JD-2450000.5) 	& 		&\multicolumn{1}{c}{($\%$)}				&\multicolumn{1}{c}{($\%$)}		&\multicolumn{1}{c}{($\%$)}	&\multicolumn{1}{c}{($\%$)} &\multicolumn{1}{c}{(deg)}\\
\hline
Feige 110			&6576.9377		&$Z$	&$0.02\pm0.08$	&$0.16\pm0.08$	&0.16$\,\pm\,$0.12	&0.11$\,\pm\,$0.12 &$-$	\\
				&6576.9227		&$J$		&$0.06\pm0.04$	&$-0.09\pm0.06$	&0.11$\,\pm\,$0.07	&0.08$\,\pm\,$0.07 &$-$	\\
				&6576.9270		&$H$	&$-0.13\pm0.07$	&$-0.02\pm0.06$	&0.13$\,\pm\,$0.10	&0.08$\,\pm\,$0.10 &$-$	\\
				&6579.0252		&$Z$	&$-0.06\pm0.14$	&$-0.08\pm0.07$	&0.10$\,\pm\,$0.16	&$\,\le\,$0.16 &$-$	\\
				&6579.0118		&$J$		&$0.03\pm0.05$	&$-0.11\pm0.06$ 	&0.11$\,\pm\,$0.08	&0.08$\,\pm\,$0.08 &$-$	\\
				&6579.0017		&$H$	&$-0.06\pm0.06$	&$-0.04\pm0.08$	&0.07$\,\pm\,$0.11	&$\,\le\,$0.11 &$-$	\\
				&6579.0055		&$K_{\rm s}^{\rm e}$	&$0.08\pm0.12$	&$-0.05\pm0.12$	&0.10$\,\pm\,$0.17	&$\,\le\,$0.17 &$-$\\
BD$+28$\,4211	&6932.8283		&$Y$	&$0.03\pm0.03$	&$-0.03\pm0.06$	&0.04$\,\pm\,$0.07	&$\,\le\,$0.07 &$-$	\\
                         	&6932.8344		&$J$		&$-0.09\pm0.06$	&$0.08\pm0.08$	&0.12$\,\pm\,$0.11	&0.06$\,\pm\,$0.11 &$-$	\\
WD 0310$-$688	&6163.4199		&$I$		&$-0.01\pm0.03$	&$-0.00\pm0.02$	&0.01$\,\pm\,$0.04		&$\,\le\,$0.04 &$-$	\\
WD 0501$+$527	&6932.8310		&$R^{\rm b}$		&$-0.02\pm0.16$	&$0.12\pm0.16$	&0.12$\,\pm\,$0.23		&$\,\le\,$0.23 &$-$	\\
                          	&6932.8321		&$R^{\rm c}$		&$-0.01\pm0.02$	&$0.02\pm0.08$	&0.02$\,\pm\,$0.08		&$\,\le\,$0.08 &$-$	\\
                          	&6932.8432		&$I^{\rm b}$		&$0.02\pm0.20$	&$0.07\pm0.10$	&0.07$\,\pm\,$0.22		&$\,\le\,$0.22 &$-$	\\
                          	&6932.8447		&$I^{\rm c}$		&$0.09\pm0.04$	&$0.02\pm0.05$	&0.09$\,\pm\,$0.07		&0.06$\,\pm\,$0.07 &$-$	\\
\hline
\multicolumn{8}{c}{{Polarized standard stars}}\\
\hline
Object			& Obs$.$ date				&Filter	 &\multicolumn{1}{c}{$q$}				&\multicolumn{1}{c}{$u$}		&\multicolumn{1}{c}{$p^{*}$}	&\multicolumn{1}{c}{$\Theta$}	&$P_{\rm lit},\,\Theta^{\rm a}$	\\
				&(JD-2450000.5)		 			& 		&\multicolumn{1}{c}{($\%$)}				&\multicolumn{1}{c}{($\%$)}		&\multicolumn{1}{c}{($\%$)}	&\multicolumn{1}{c}{(deg)}		\\
\hline
Hiltner 652		&6162.9848						&$I$	       &$5.43\pm0.01$	                                       &$-0.67\pm0.01$	      			&5.48$\,\pm\,$0.04	&176.5$\,\pm\,$0.2	&5.61$\pm$0.04, 179.18$\pm$0.11\\
HD 38563c		&6576.2742		&$J$&$-4.83\pm0.02$	&$2.63\pm0.03$			&5.50$\,\pm\,$0.11	&75.7$\,\pm\,$0.5	&6.03$\pm$0.10, 71$\pm$1\\
                 		&6576.2781 		&$H$&$-2.70\pm0.05$	&$1.52\pm0.07$			&3.10$\,\pm\,$0.14	&75.4$\,\pm\,$1.3	&3.68$\pm$0.10, 70$\pm$1\\
                		&6578.2703		&$Z$&$-6.40\pm0.10$	&$3.58\pm0.06$			&7.33$\,\pm\,$0.18	&75.4$\,\pm\,$0.7	&7.27$\pm$0.10, 71$\pm$1\\
                		&6578.2737	&$J$&$-5.20\pm0.15$	&$3.11\pm0.13$			&6.06$\,\pm\,$0.22	&74.5$\,\pm\,$1.0	&6.03$\pm$0.10, 71$\pm$1\\
                		&6578.2794	&$H$&$-2.94\pm0.07$	&$1.65\pm0.06$			&3.37$\,\pm\,$0.14	&75.3$\,\pm\,$1.2	&3.68$\pm$0.10, 70$\pm$1\\
HD 283855		&6933.0791	&$Y$&$-0.44\pm0.03$	&$2.89\pm0.05$			&2.92$\,\pm\,$0.12	&49.3$\,\pm\,$1.2	&3.27$\pm$0.03, 46$\pm$1\\
                   		&6933.0842	&$J$&$-0.24\pm0.09$	&$2.29\pm0.02$			&2.30$\,\pm\,$0.13	&48.0$\,\pm\,$1.7	&2.58$\pm$0.03, 46$\pm$1\\
BD$+$64\,106       &6933.0971	&$R^{\rm b}$&$-5.04\pm0.15$	&$-0.78\pm0.14$			&5.09$\,\pm\,$0.30	&94.4$\,\pm\,$1.7	&5.15$\pm$0.10, 96.7$\pm$0.5\\
                              &6933.0942	 &$I^{\rm b}$&$-4.63\pm0.10$	&$-0.77\pm0.18$			&4.69$\,\pm\,$0.30	&94.7$\,\pm\,$1.8	&4.69$\pm$0.05, 96.9$\pm$0.3\\
                              &6933.0980	&$R^{\rm c}$&$-5.08\pm0.07$	&$-0.86\pm0.06$			&5.16$\,\pm\,$0.12	&94.8$\,\pm\,$0.7	&5.15$\pm$0.10, 96.7$\pm$0.5\\
                              &6933.0952 	&$I^{\rm c}$&$-4.60\pm0.06$	&$-0.72\pm0.08$			&4.65$\,\pm\,$0.12	&94.4$\,\pm\,$0.7	&4.69$\pm$0.05, 96.9$\pm$0.3\\

\hline\hline
\end{tabular}
\begin{minipage}{175.5mm}
Notes: $^{\rm a}$ Literature data from ALFOSC manual on http://www.not.iac.es/instruments/alfosc/polarimetry/index.html, \citet{1992ApJ...386..562W}, and \citet{2007ASPC..364..503F}. $^{\rm b}$ Average of the four measurements using four retarder plate angles. $^{\rm c}$ Measurement obtained from the total of 16 retarder plate angles (see text).
\end{minipage}
\end{table*}


\section{Taurus objects}\label{tau}
We discuss CFHT-BD-Tau 4 and KPNO-Tau 4 separately from the rest of the target list since these two objects are members of the Taurus star-forming region and have a very distinct age of $\approx$1 Myr. Both are affected by moderate amounts of extinction: $A_{\rm v}\sim$ 2.5, $A_{J}\sim0.9$ mag (KPNO-Tau 4) and $A_{\rm v}\sim$3, $A_{J}\sim1.9$ mag (CFHT-BD-Tau 4) according to \citet{2007A&A...465..855G}, part of which might come from the presence of surrounding disks responsible for the infrared flux excesses \citep{2010ApJS..186..111L}. Additionally, several works hint to the presence of accretion events \citep{2003ApJ...592..282J,2005ApJ...626..498M,2005ApJ...625..906M}. All of these properties clearly differentiate the Taurus sources from the field sample in relation to polarimetric studies.  

CFHT-BD-Tau 4, the warmest and most massive of the two Taurus brown dwarfs in our analysis, has time consecutive $Z$- and $J$-band photometry, separated by 25.2 min, displaying significant level of linear polarization. The polarization degree and vibration angle measurements are compatible within 1-$\sigma$ the quoted uncertainties, and neither $p^{*}$ nor $\Theta$ appear to show any dependency on wavelength. KPNO-Tau 4, which has a mass near the brown dwarf---planet borderline at the age of Taurus, was observed in the $J$-band on three different occasions spread over $\approx$1 yr. Interestingly, significant linear polarization was found on two distinct moments, and although the degrees of polarization coincide within 1-$\sigma$ the error bars, the vibration angles deviate from one measurement to the next. This indicates strong variability in the $J$-band linear polarization angle of KPNO-Tau 4.

The origin of the observed linear polarization in CFHT-BD-Tau 4 and KPNO-Tau 4 can be explained by a few scenarios that may occur independently or all at once. One is the interstellar extinction towards the Taurus region (Taurus objects are located at 140 pc). \citet{1993A&A...274..203L} and \citet{2002A&A...394..675T} argued that the effects of the interstellar dust on linear polarization become noticeable typically at distances $\ge$\,70 pc. This kind of linear polarization remains constant on short and long time scales. Therefore, it cannot account for the polarimetric variability seen in KPNO-Tau 4. One second possible explanation is the internal extinction of Taurus due to the presence of the parental clouds remnants. The third and most likely picture is a polarization caused by the presence of disks of dust and gas surrounding both objects. In this scenario, variability on scales of days to months is expected as a consequence of the rotation of a inhomogeneous disk around the central object and/or a rotating spotted central source illuminating the disk (e.g., \citealt{2003A&A...409..163M}). However, \citet{1981ApJ...247.1013P} found that the polarimetric measurements of two spotted flare stars showed no variability in polarization when starspot groups move across the stellar disk with rotation. This leaves the disk scenario as the most plausible explanation. It might be also feasible that the polarimetric signal is originated by the scattering of light due to dusty clouds intrinsic to the brown dwarfs. We note that the linear polarimetry degrees of CFHT-BD-Tau 4 and KPNO-Tau 4 do not differ significantly from those found in field ultra-cool dwarfs (see next Section) despite having primordial surrounding disks.

\section{Young field dwarfs}
The remaining targets have age estimations in the range $\sim$10--500 Myr, except for J0033$-$1521, J0241--0326, and J1552$+$2948, which are thought to be older than 500 Myr (see Section~\ref{intro} and Table~\ref{Table1}). All are nearby ($\le$\,50 pc), polarization due to the interstellar extinction is thus supposed to be unimportant and we made no attempt to correct for it in our data. In the following, we explored the linear polarimetric properties of the field sample as a function of the different observing time scales, wavelength, and age.


\begin{figure*}
\centering
\includegraphics[width=17cm]{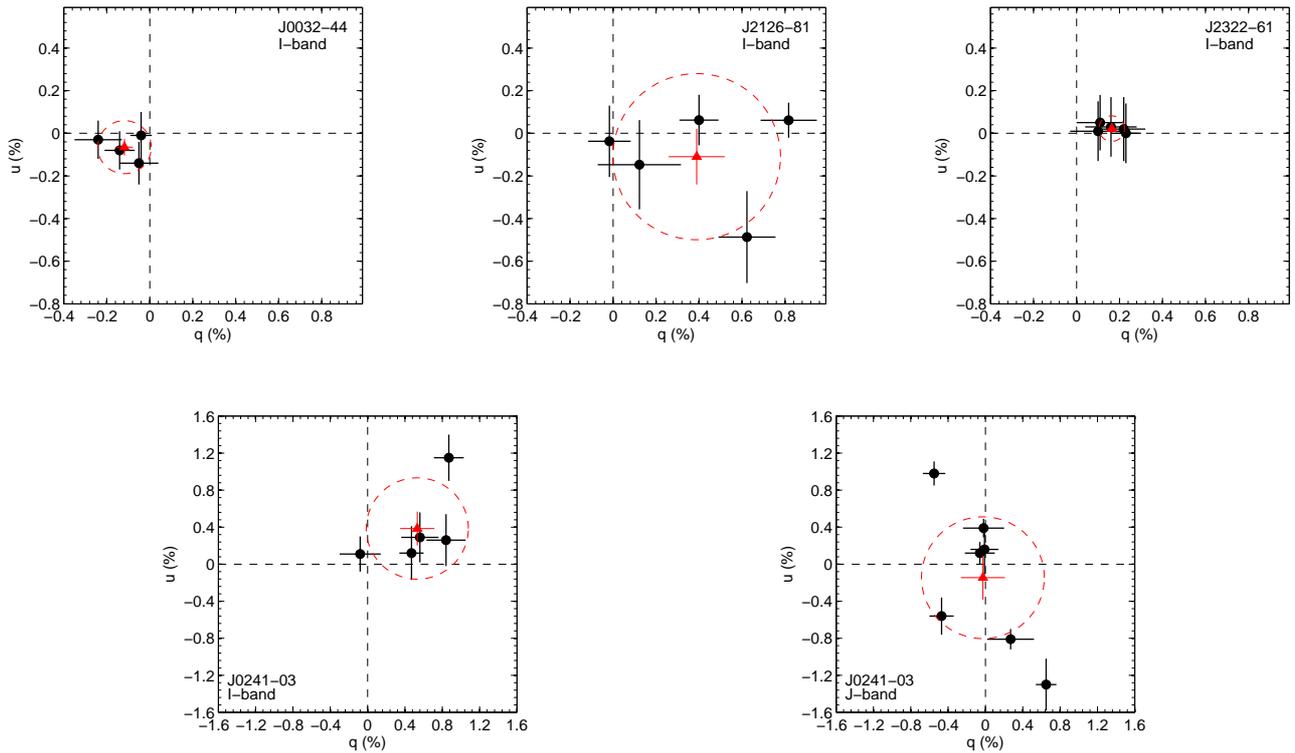}
\caption{$I$- and $J$-band Stokes $q-u$ planes for the four targets with polarimetric time series observations (J0032--4405, J2126--8140, J2322--6151, and J0241$-$0326). Individual measurements and their associated error bars are plotted as black symbols. The red triangle denotes the average $q$ and $u$ values, and the red dashed circle represents their 99\%~confidence interval following \citet{1986VA.....29...27C}.}
\label{Fig_qu}
\end{figure*}

\subsection{Polarization versus time\label{pt}}

\subsubsection{Short-term polarimetric variability (rotation)}
Four sources (J0032$-$4405, J0241$-$0326, J2126$-$8140, and J2322$-$6151) were monitored in the $I$-band with a cadence of $\sim$20$-$46 min over continuous $\sim$2.1$-$7.5 h. J0241--0326 was also monitored in the $Z$-, $J$-, and $H$-filters for a total duration of $\sim2.2$ h. In these time series, the number of polarimetric data points per target ranged from four to ten. These observations allowed us to study the polarimetric variability at short-time scales or with rotation. None of the four ultra-cool dwarfs has rotation periods or spectroscopic rotational velocities available in the literature. Nevertheless, we attempted to estimate the possible rotational periodicities of the sample by assuming a typical radius of $\le1.5$ R$_{\rm J}$ for intermediate-gravity objects and $v$\,sin\,$i\approx10-90$ km\,s$^{-1}$, which are the values measured for field dwarfs of similar spectral type at both young \citep[$\sim$1--10 Myr; ][]{2005ApJ...626..498M} and mature ages \citep{2008ApJ...684.1390R,2010ApJ...723..684B}. We found that the rotation periods of intermediate-age ultra-cool dwarfs could be in the interval $\sim2-18$ h as seen in Pleiades and  alpha Persei low mass members \citep{1999AJ....118.1814T,1997MNRAS.286L..17M}. Therefore, our time series of $I$-band polarimetric data likely covered a significant fraction of a rotation or one complete rotation per object.

The normalized Stokes $q$ and $u$ data of the $I$- and $J$-band time series are plotted in Figure~\ref{Fig_qu}, one ultra-cool dwarf per panel. We also plotted the average $q$ and $u$ values and computed their 99\%~confidence interval following \citet{1986VA.....29...27C}, which was used as a test for the presence of polarimetric variability. From Figure~\ref{Fig_qu}, J0032$-$4405 and J2322$-$6151 show Stokes parameters consistent within 1-$\sigma$ the quoted error bars and located very near the 99\%~confidence circle, indicating that no $I$-band linear polarimetric variability with an amplitude higher than $\sim$0.1--0.2\%~is detected in any of these two dwarfs. Despite having $q$ and $u$ uncertainties and a number of data points similar to  J0032$-$4405 and J2322$-$6151 (Table~\ref{Table2}), J0241$-$0326 and J2126$-$8140 display a larger dispersion in Figure~\ref{Fig_qu}. The former object has some $I$-and $J$-band measurements out of the 99\%~confidence circle in the $q-u$ planes, which might indicate some polarimetric variability of notorious amplitude. 

\begin{figure*}
\centering
\includegraphics[width=17cm]{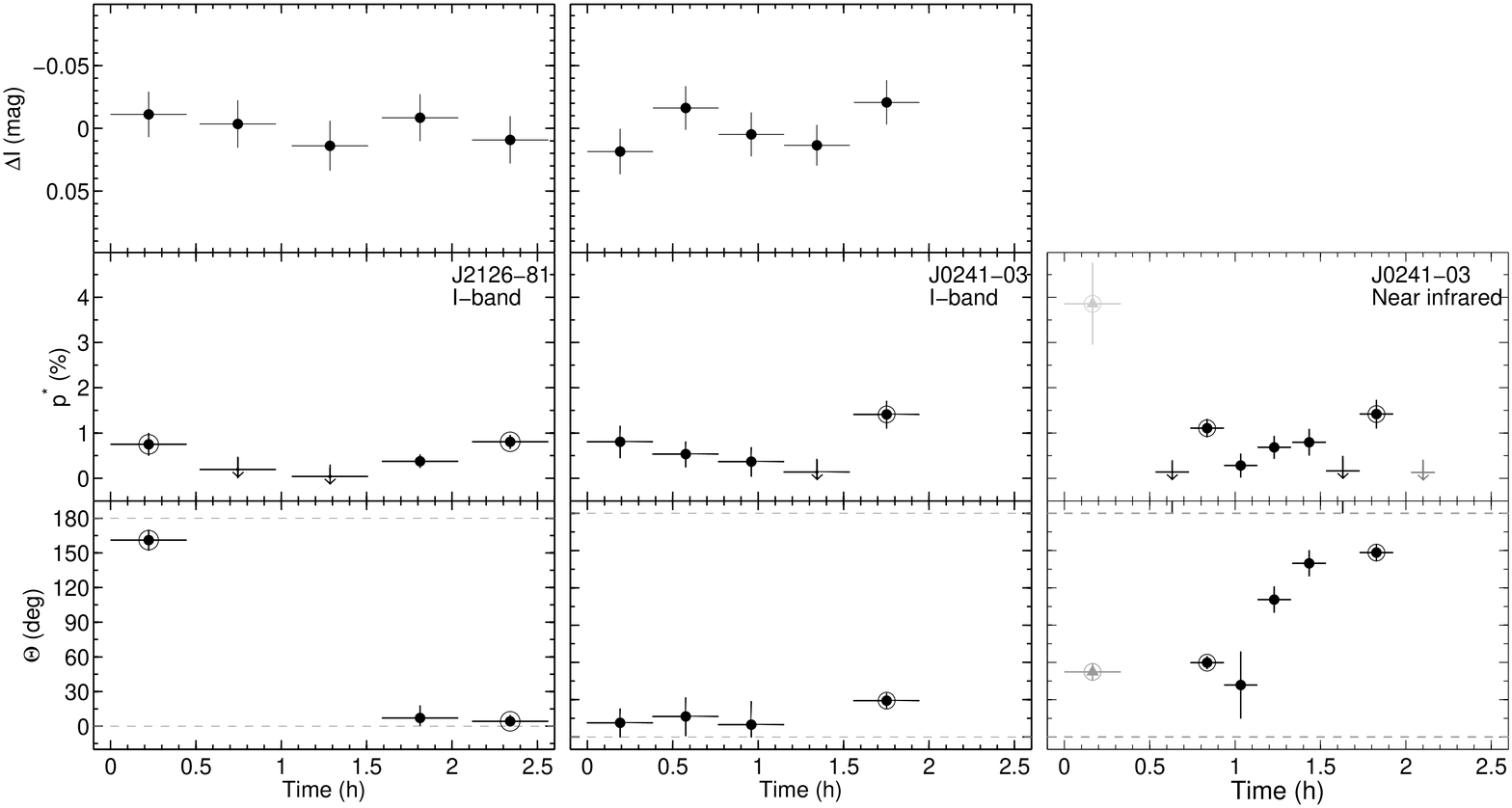}
\caption{Photometric linear polarimetry time series ($p^{*}$, $\Theta$) of J2126--8140 {\sl (left)} and J0241--0326 {\sl (middle, right)}.  The intensity light curves are also shown for the optical data on the two top panels ({\sl left} and {\sl middle}). The unbiased degree and the vibration angle of the polarization are depicted on the middle and bottom panels, respectively. Zero times correspond to Julian Dates of 2456163.4953 (J2126--8140, {\sl left}), 2456163.7062 (J0241--0326, $I$, {\sl middle}), and 2456579.5487 (J0241--0326, $Z$, {\sl right}). The symbols are plotted at the mid points of the polarimetric cycles, vertical error bars stand for the measurements uncertainties, and the horizontal errors account for the duration of the polarimetric cycles. Down arrows indicate 1-$\sigma$ upper limits. Data compliant with $P/\sigma \ge 3$ are shown with encircled symbols. In the {\sl right} panel, $Z$-, $J-$, and $H$-band photometry is plotted as triangles, dots, and squares, respectively. We caution that the error bars associated with $\Theta$ ($1 < P/\sigma_p < 3)$ may be underestimated.}
\label{Fig3}
\end{figure*}

Figure~\ref{Fig3} depicts the $p^*$ and $\Theta$ time series observations of J0241$-$0326 ($I$, $J$) and J2126$-$8140 ($I$). The angle $\Theta$ was not computed for those values with $\sigma_P \ge P$. The $I$-band degree of linear polarization of J2126$-$8140 decreases and increases with a peak-to-peak of $\sim$0.8\%~in a time scale of about 2 h (left panel of Figure~\ref{Fig3}). The polarization vibration angle changes from $161\pm9$ deg to small values ($2\pm6$ deg) within 2 h (or equivalently from $-19\pm9$ to $2\pm6$ deg as also seen in Fig. \ref{Fig_qu}). The middle and right panels of Figure~\ref{Fig3} correspond to the $I$- and $J$-band data of J0241$-$0326. The two sets of observations are not simultaneous but taken $\sim1.1$ yr apart. In the $I$-band, there is a swallow decrease of the linear polarization degree from $\sim$0.9\% to $\le$0.24 \% during $\sim$1.5 h, and a sudden increase to $p^* \sim 1.4\%$ (with a significance of $\sim$4.7-$\sigma$) that occurs in about 20 min. However, $\Theta$ remains nearly unchanged between $10^{\circ}$ and $30^{\circ}$ (i.e., typically within 1.5-$\sigma$ the associated error bars). The $J$-band polarimetric light curve of J0241$-$0326 has a different (more stochastic) pattern, yet the measured $I$- and $J$-band linear polarization degrees are quite alike, a fact that agrees with the predictions by \citet{2010ApJ...722L.142S}. The $J$-band polarization vibration angle shows a progressive change by about $88\fdg6$ in the interval of $\sim$ 1 h. 

By combining the ordinary and extraordinary rays of the FORS2 polarimetric images, we built the differential $I$-band intensity light curves for J0241$-$0326 and J2126$-$8140 using reference stars of brightness similar to our targets. There are no sufficient bright reference stars in the field of the $J$-band data of J0241$-$0326 to obtain this object's near-infrared intensity light curve. The procedure to derive the intensity light curves from the polarimetric data is described in \cite{2015A&A...580L..12M}. The resulting differential $I$-band light curves, shown in the top panels of Figure~\ref{Fig3}, have photometric dispersions of $\pm$16 mmag (J0241$-$0326) and $\pm$10 mmag (J2126$-$8140), which basically coincide with the estimated protometric error bars associated with the individual data points. This suggests that no obvious variability larger than a few tens of mmag is seen in the $I$-band emission of these two targets, which contrasts with the polarimetric curves. It may be possible that small changes in the atmospheric dust distribution/concentration have large impact on the linear polarization while remaining imperceptible (or below a few tens of mmag) in the intensity light curves.

 J2208$+$2921 has a rotation period of 3.5\,$\pm$\,0.2 h \citep{2015ApJ...799..154M} obtained from sinusoid-like {\sl Spitzer} $[3.6]$ and $[4.5]$ light curves with broad amplitudes of $\sim$69 and $\sim$ 54 mmag, respectively, consistent with its young age. There is only one $J$-band linear polarimetric measurement for this L3 dwarf indicating a low level of light polarization. More polarimetric data combined with intensity light curves at various filters may provide new insights into the origin of the observed variability.

Under the scenario of linear polarization being caused by the presence of dusty atmospheric upper layers (both J0241$-$0326 and J2126$-$8140 have early-L spectral types and are relatively warm), the data of Figure~\ref{Fig3} suggest that regardless of the shape of the sources there are structures in the atmospheres revolving with the dwarfs. Oblateness and a homogeneous dusty atmosphere produce a constant polarization throughout rotation; therefore, they alone cannot account for polarimetric changes on time scales of one rotating cycle. The atmospheric structures may be composed of increased concentrations of dust (``clouds"), or holes in a dusty atmosphere (where deeper and dust-free atmospheric layers can be seen).

It is now well established that ultra-cool dwarfs with late-M and early-L types can host strong magnetic fields \citep[e.g., ][]{2006ApJ...648..629B,2012ApJ...746...23M,2014ApJ...785....9W}. They may cause linear polarization; however, it is expected to be small at optical and near-infrared wavelengths. \citet{2002A&A...396L..35M} and \citet{2004sf2a.conf..305M} studied the linear polarization properties of M dwarfs, some of which are known to have significant magnetic fields; yet, their polarization degrees are typically below $P\,\sim$ 0.2\% in the $I$-band. The L3.5 dwarf 2MASS J00361617$+$1821104 has an associated variable and periodic radio emission consistent with a large-scale magnetic field of about 175 G \citep{2006ApJ...648..629B}. \citet{2013A&A...556A.125M} measured a linear polarization degree of $p^*$\,=\,0.20$\,\pm\,$0.11 \%  in the $J$-band for this particular source, and discussed that the likely contribution of the magnetic field to polarization in the near-infrared is very small provided that gyro- and synchrotron processes are associated with the field.

\subsubsection{Long-term polarimetric variability}
According to the theory of ultra-cool dwarf atmospheres and the high rotation observationally inferred, the upper atmospheric layers are expected to be highly dynamical, including significant circulation at regional and global scales. Consequently, any inhomogeneous feature within the atmosphere may evolve significantly after a few to several rotations (e.g., \citealt{2010A&A...513A..19F,2013ApJ...776...85S,2014Natur.505..654C}). This may induce polarimetric variability on long time scales. The field targets J0045$+$1634, J0241$-$0326, J1552$+$2948, and J1726$+$1538 were observed with the same instrumental configuration on different occasions separated by months to years. They allowed us to address the long-term polarimetric variability.

\begin{figure}
   \centering
\includegraphics[width=0.49\textwidth]{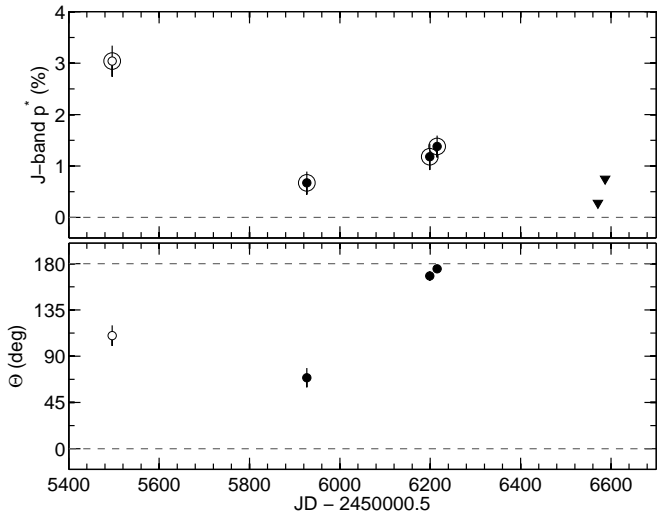}
     \caption{LIRIS $J$-band linear polarimetry of J0241-0326 on long-term scales. The polarization degree and the polarization vibration angle are shown on the top and bottom panels. The open circle stands for \citet{2011ApJ...740....4Z} datum. Data corresponding to 2012 October 6--7 and 2013 October 12--13 are slightly shifted for clarity. Down-looking triangles represent 1-$\sigma$ upper limits. The most recent measurement represents the amplitude of the time series data of Figure~\ref{Fig_qu}. Encircled symbols as in previous Figures.}
              \label{Fig5}%
    \end{figure}

In our sample, J0241$-$0326 has the longest time coverage spanning 2.95 yr and the largest number of observations. This L0 dwarf was first reported to be linearly polarized ($p^{*}=3.04\pm0.30\%, \Theta=110\pm10$ deg, $J$-band) by \citet{2011ApJ...740....4Z}. These authors also used LIRIS and identical instrumental configuration. Therefore, our and their data can be safely compared. J0241$-$0326 has been monitored on five different occasions since then (see Table~\ref{Table2}). Figure~\ref{Fig5} depicts all six $J$-band polarimetric photometry ($p^*$ and $\Theta$) available for the L0 dwarf as a function of observing time. For simplicity, the time series observations discussed in the previous Section were averaged (one data point per observing night) and the dispersion of the measurements associated with this epoch is shown as a down-looking triangle in Figure~\ref{Fig5} to illustrate the amplitude of the short-term polarimetric light curves. The averaged $p^{*}$ and $\Theta$ values taken on consecutive nights are consistent within 1-$\sigma$ the uncertainty, suggesting that there is little polarimetric variability from night to night ($\Delta p^* \le 0.25\%$), this is after a low to moderate number of rotations. This result agrees with the findings for high-gravity dwarfs of similar spectral classification by \citet{2013A&A...556A.125M}. The linear polarimetric behavior of J0241$-$0326 has a different pattern on longer time scales (e.g., months): The changes in both the polarization intensity and vibration angle are significant with peak-to-peak variations as high as $\sim3\%$. Interestingly, \citet{2014A&A...568A..87L} reported a peak-to-peak photometric variability of $44\pm10$ mmag in the long-term intensity of the $K$-band emission with a tentative periodicity of $\sim300$ d. The changes in the polarimetric data of J0241$-$0326 agree with this time scale.

J0045$+$1634 has three $J$-band polarimetric measurements including the two reported here and the very first one ($p^{*}_{J}=\,0.00\,\pm\,0.11\,\%$) obtained by \citet{2013A&A...556A.125M}. All three data were collected with the same instrumental setup and in intervals of approximately 1 yr (the second and third measurements were taken 1.01 and 1.98 yr after the first observation). The $J$-band linear polarimetry of this L2 dwarf changes  between $\le$0.11\% and $0.36\pm0.12\%$, which indicates some variability at the 2--3-$\sigma$ level. The case of J1552$+$2948 is similar to that of J0045$+$1634. Of the three $J$-band measurements available for J1552$+$2948 that are equally spaced in time and cover a total of 1.08 yr, only one has a significant degree of polarization with values ranging from $\le$0.12\% to $0.45\pm0.15\%$. \citet{2010ApJ...723..684B} determined the spectroscopic rotational velocities of J0045$+$1634 and J1552$+$2948 to be $v$\,sin\,$i = 32.8 \pm 0.2$ km\,s$^{-1}$ and $ 18.9 \pm 0.6$ km\,s$^{-1}$; and  \citet{2014A&A...568A...6Z} provided T$_{\rm eff}\sim 1970, 2260$ K and gravities of log\,$g$ = 4.5--4.66, $\ge$5.13 [cm\,s$^{-2}$], respectively. Despite their likely differing surface gravities, the amplitude of the long-term polarimetric observations is similar for both the L0 and L2 dwarfs.

J1726$+$1538 was observed on two occasions separated by $\sim$7.5 months. Both measurements show high levels of $J$-band linear polarization compatible at the 1-$\sigma$ the quoted uncertainties. The vibration angle of the polarization differs notably between the two epochs, suggesting some polarimetric variability. \citet{2014A&A...568A..87L} reported no obvious photometric, long-term variability in the intensity of the $K$-band emission of J0045$+$1634, J1552$+$2948, and J1726$+$1538, with upper limits on the peak-to-peak variations of about 35 mmag. The short-term, non-variability nature of the intensity light curves of J1726$+$1538 was also confirmed by \citet{2015ApJ...799..154M}, who reported upper limits of $\sim$29 and $\sim$49 mmag on the amplitudes of the {\sl Spitzer} $[3.6]$ and $[4.5]$ time series data

J0355$+$1133 has $J$-band linear polarimetry data published in \citet{2011ApJ...740....4Z}, which were taken 3 yr prior to the present work. None of the measurements indicate strong polarization; however, the first epoch observation has an error bar about four times greater than the one quoted here.

From our data, we concluded that nearly all field, early-L dwarfs with polarimetric observations spread over long time scales display variable $J$-band linear polarimetry, even though they do not necessarily show significant long-term variability of the intensity light at other infrared wavelengths. This contrasts with the short-term $I$-band data, where only half of the sample appears to be variable in time scales typically within a rotation. The amplitudes of the short- and long-term variability of the linear polarimetry degree are alike. Dusty atmospheres with evolving structures on varying time scales provide a feasible explanation for these observations. Additional data with better time samplings would be required to study whether there are cycles of ``atmospheric patterns" at all temporal scales.

\subsection{Linear polarization versus age \label{age}}
Despite any possible linear polarization variability, it is worth studying whether there is a trend of the intensity of the linear polarization at near-infrared wavelengths with age. This might provide new insights on the amounts of dust in the atmospheres depending on the surface gravity. 

\begin{figure}
\centering
\includegraphics[width=0.49\textwidth]{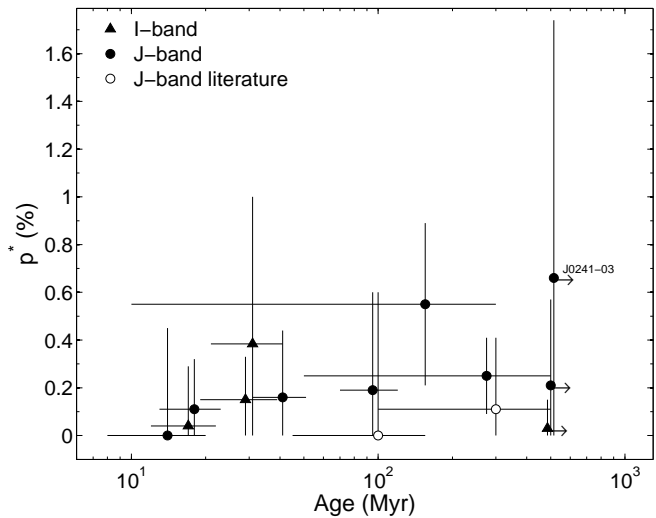}
\caption{Unbiased degree of linear polarization as a function of age. For simplicity, each object is represented by one average degree of polarization with a vertical error bar that accounts for the spread of the polarimetric observations. The horizontal errors stand for the age intervals. Age lower limits are plotted as arrows. Objects with the same age are slightly shifted for clarity. The data of this paper are plotted as filled symbols, and data from the literature as open circles.}
\label{Fig6}
\end{figure}

Figure~\ref{Fig6} depicts the unbiased $I$- and $J$-bands linear polarization degree as a function of age (according to Section~\ref{targets}) for all objects in our sample minus the Taurus sources. Both filters are shown altogether because the polarimetric values are indistinguishable given the uncertainties. We also collected the $J$-band data of young dwarfs (G\,196$-$3\,B, J0045$+$1634, J0355$+$1133, and 2MASS\,J10224821$+$5825453) from \citet{2011ApJ...740....4Z} and \citet{2013A&A...556A.125M} for completeness. For each dwarf, we plotted an averaged degree of polarization with associated error bars that account for the observed spread (including their individual uncertainties). The data of Figure~\ref{Fig6} present a significant scatter ($p^*$\,$\approx$\,0--1\%) from intermediate to old ages; there is no obvious pattern beyond the quoted errors. This is likely due to the limited number of data (small statistics) and to the observed polarimetric variability, which significantly contributes to the blurring of the Figure. Additionally, any error on the age determination can lead to inconclusive results. For example, the L dwarf with the largest variations in our target list, J0241$-$0326 (labeled in Figure~\ref{Fig6}), has a contradictory age according to different works (see Table~\ref{Table1}). Based on its location in the Hertzsprung-Russell diagram and the lithium non-detection in its optical spectrum (suggesting lithium depletion), \citet{2014A&A...568A...6Z} argued that J0241$-$0326 is likely older than a few hundred Myr, whereas \citet{2013ApJ...772...79A} discussed that this L0 dwarf has a triangular-shaped $H$-band continuum (typical of low gravity atmospheres), and \citet{2014ApJ...783..121G} considered it a likely member of the $\sim$30 Myr Tucana-Horologium moving group. 

Regardless of the location of J0241$-$0326 in Figure~\ref{Fig6}, the remaining polarimetric data do not delineate a clear trend with age. If any, the youngest dwarfs in the sample (J0032$-$4405, PSO\,J318, and J2208$+$2921) show small amounts of linear polarization. According to the theory \citep{2001ApJ...556..357A,2011ApJ...733...65B,2011ApJ...735L..39B}, low gravity dwarfs are expected to have dustier atmospheres than high gravity dwarfs. This implies that more multi-scattering processes occur in random planes of the atmospheres, cancelling their contribution to the polarization of the light and resulting in low degrees of linear polarization \citep{2003ApJ...585L.155S}. This might provide a feasible scenario for the small polarization degrees of the reddest (therefore, likely dustier) dwarfs in our sample (J0355$+$1133 and PSO J318$-$22). On the contrary, high gravity dwarfs may have less atmospheric dust producing single scattering events and high values of polarization. Unfortunately, our data are not sufficient to robustly confirm or discard this theoretical prediction.

   \begin{figure}
    \resizebox{\hsize}{!}{\includegraphics{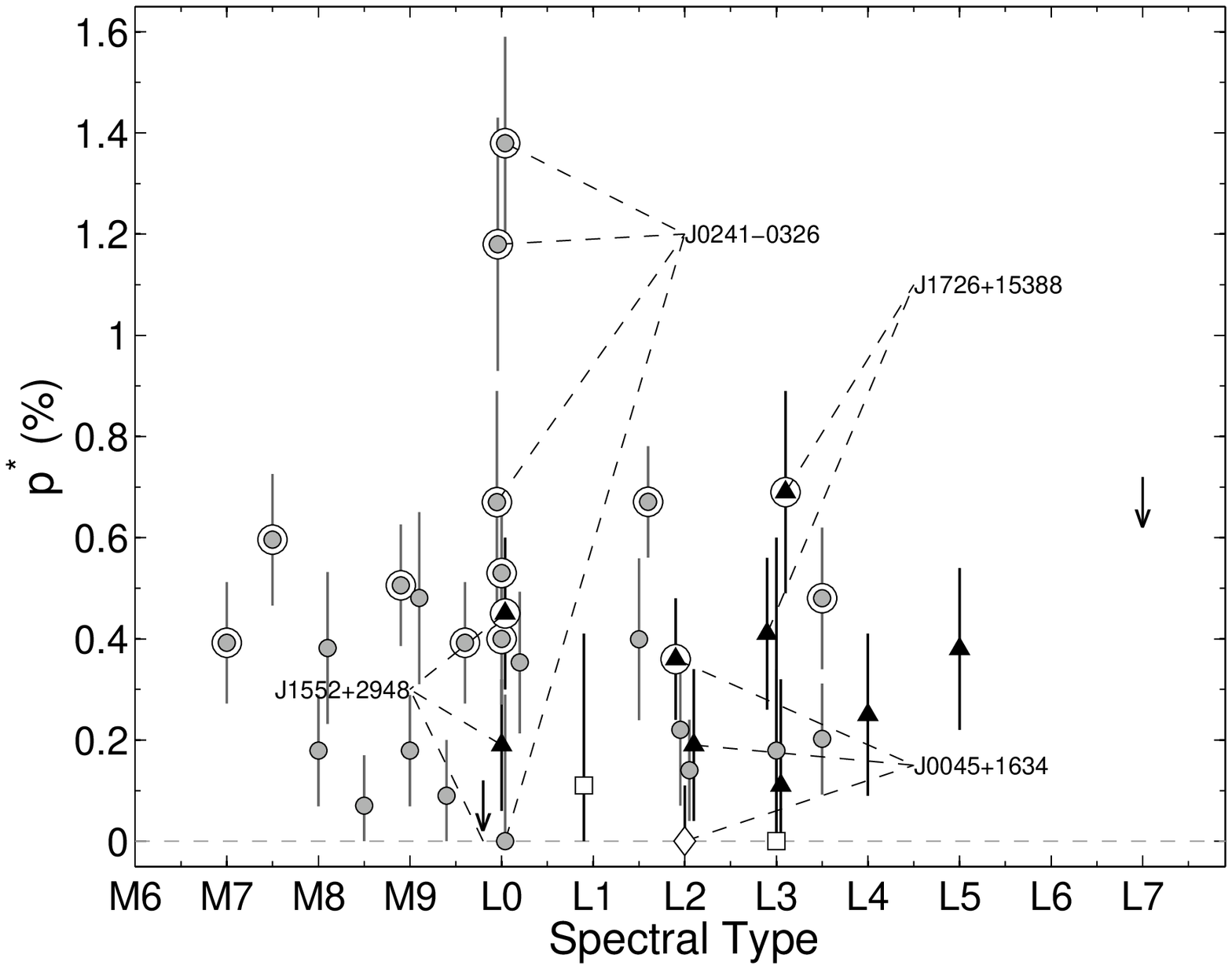}}
     \caption{$J$-band unbiased linear polarization degree as a function of late-M and L spectral types. Young dwarfs are plotted as filled black triangles, likely older dwarfs ($\ge0.5$ Gyr) taken from \citet{2013A&A...556A.125M} are shown with gray circles. Data from \citet{2011ApJ...740....4Z} are depicted with white symbols. Detections at the level of 3-$\sigma$ are encircled. Down arrows indicate 1-$\sigma$ upper limits. The typical uncertainty in the spectral classification is half a subtype (not plotted for clarity). Objects with several polarimetric measurements are labeled and slightly shifted in spectral type.}
              \label{Fig7}%
    \end{figure}

Our $J$-band measurements of young ultra-cool dwarfs and the polarimetric data obtained for field, high-gravity, fast-rotating dwarfs of related classification \citep{2013A&A...556A.125M} are depicted as a function of spectral type in Figure~\ref{Fig7}. The two data sets have similar polarimetric precisions ($\pm$0.10--0.25\,\%) and were constructed using the same instrumental configuration; consequently, they can be compared. From Figure~\ref{Fig7}, the polarimetric indices of both samples show significant scatter and typically lie below $p^* \approx 2\%$. Based on current data, there are no apparent differences between the two sets that could be ascribed to divergent surface gravities. Polarimetric data obtained to a higher precision and a systematic monitoring to analyze the amplitude of the polarimetric variations would be required to search for any possible deviation. 

\subsection{Linear polarization versus wavelength\label{wav}}
The degree of linear polarization has its maximum value at wavelengths near the typical size of the grains responsible for the light scattering (e.g., \citealt{2001ApJ...561L.123S}). Therefore, a multiwavelength study of the linear polarimetric properties of ultra-cool dwarfs may provide critical insight onto the sizes of the atmospheric particles. Because of variability (see Section~\ref{pt}), low-resolution spectropolarimetry or simultaneous photometric observations or observations taken very close in time using different filters are demanded. In our study, the Taurus dwarf CFHT-BD-Tau 4, and the dwarfs J0241$-$0326 and J0045$+$1634 meet this requirement. 

\begin{figure}
\centering
\includegraphics[width=0.49\textwidth]{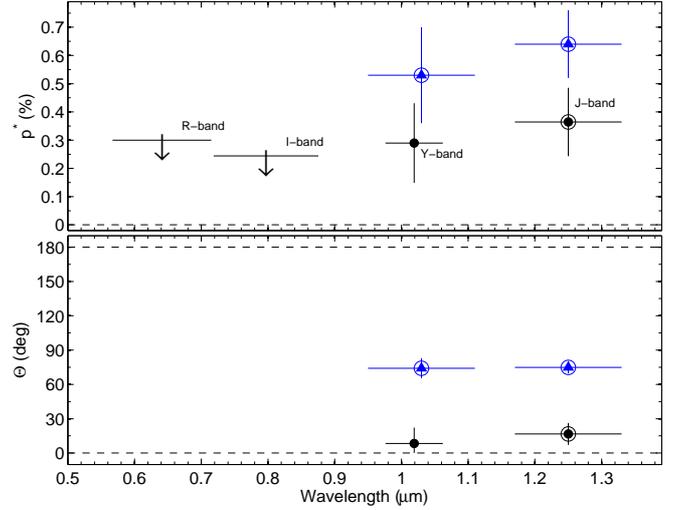}
\caption{Linear polarimetry of J0045$+$1634 (black circles) and CFHT-BD-Tau 4 (blue triangles) as a function of wavelength (simultaneous data). Vertical bars stand for errors in the polarization degree (top panel) and the polarization vibration angle (bottom panel); horizontal bars account for the filter widths. Upper limits on polarization are indicated with an arrow. The 3-$\sigma$ criterion compliant data have encircled symbols.}
\label{Fig9}
\end{figure}

On 2014 October 2 we obtained the simultaneous $RIYJ$ degrees of linear polarization of J0045$+$1634, thus covering from $\sim$0.6 through 1.2 $\mu$m. The measurements are depicted in Figure~\ref{Fig9}. The large error bars associated with the optical data prevented us from concluding on the wavelength dependency of these polarimetric observations. At the 1-$\sigma$ level, the $Y$- and $J$-band data have compatible linear polarization degree and vibration angle. We also included CFHT-BD-Tau 4 (discussed in Section \ref{tau}) in Figure \ref{Fig9}.

\begin{figure}
\centering
\includegraphics[width=0.49\textwidth]{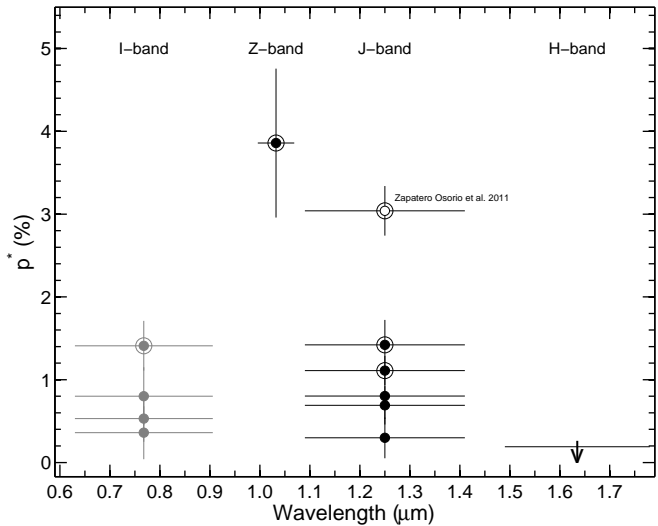}
\caption{Linear polarimetry of J0241$-$0326 as a function of wavelength. The $Z$-, $J$-, and $H$-band data (black symbols) were taken consecutively within a time interval of 2 h. Other non-simultaneous observations are shown with white and gray symbols. Error bars as in Figure~\ref{Fig9}. The 3-$\sigma$ criterion compliant data have encircled symbols.}
\label{Fig8}
\end{figure}

We measured the linear polarization of J0241$-$0326 in three different bands ($Z$, $J$, and $H$) within a time interval of $\sim2$ h on 2013 October 13. The $J$-band time series was presented in Figure~\ref{Fig3} and discussed in Section~\ref{pt}. Figure~\ref{Fig8} illustrates all consecutive $ZJH$ polarimetric measures; it shows that the largest degree happens at the shortest wavelength ($\sim$1.03 $\mu$m) and that the linear polarization intensity decreases towards long wavelengths. Even though the $J$-band data present variability, its maximum amplitude lies below the single $Z$-band measurement. Furthermore, the large $J$-band polarization degree found by \citet{2011ApJ...740....4Z}, also plotted in Figure~\ref{Fig8} for a proper comparison, is surpassed by the $Z$-band data. \citet{2013ApJ...776...85S} predicted that the dust clouds of ultra-cool dwarfs could be mainly built up by particles smaller than 1 $\mu$m. More recently, \citet{2016ApJ...830...96H} assumed the existence of a ``dust haze'' of small particles in the upper atmospheres of red L dwarfs, and found that sub-micron-range silicate grains with mean effective radius between 0.15 and 0.40 $\mu$m can reproduce the observed spectral energy distributions of red L dwarfs. It is thus expected that the largest values of $p^*$ occur at wavelengths shorter than about 1 $\mu$m. The trend delineated by our simultaneous $ZJH$ polarimetric observations of J0241$-$0326 supports \citet{2016ApJ...830...96H} finding for this particular source. The $I$-band data of J0241$-$0326 (also shown in Figure~\ref{Fig8}) were not simultaneous with the $ZJH$ measurements. Yet, the $I$-band polarimetric amplitude is similar to that of the $J$ filter and smaller than the $Z$-band, which might indicate that the linear polarization is maximum at around 1 $\mu$m in J0241$-$0326. Further data are required for a robust confirmation.

\section{Conclusions \label{conclusions}}

We used FORS2 on the Very Large Telescope, ALFOSC on the Nordic Optical telescope, and LIRIS on the William Herschel telescope to measure optical ($RIZ$) and near-infrared ($YJHK_{\rm s}$) broadband linear polarization in a sample of 12 M7--L7 field low-gravity dwarfs and two Taurus brown dwarfs. All have likely ages ranging from $\approx 1$ through $\approx$ 500 Myr and show spectroscopic signatures of low- and intermediate-gravity and have very red colors, which are thought to be produced by dusty atmospheres (field young dwarfs) or disks (Taurus dwarfs). This provides an scenario where we can investigate the linear polarization signal.

We achieved linear polarization accuracies of $\sigma_{P}=\,\pm\,0.1-0.9\%$ depending on the observing filter and brightness of the targets. The two Taurus brown dwarfs appear to be polarized with degrees of $0.4-0.9 \%$ in the $J$-band. The origin of the observed polarization may likely reside in the light scattering within the disks, which are known to exist around both objects. For one of the Taurus brown dwarfs, we collected quasi-simultaneous $Y$- and $J$-band linear polarimetric photometry, finding similar polarization degrees and angles within 1-$\sigma$ for the two wavelengths. As for the field dwarfs in our sample, the degrees of linear polarization were typically measured in the interval $\approx 0-1.5\%$, finding that the polarization intensity is similar at both the $I$- and $J$-bands, which agrees with the theoretical predictions for early L dwarfs. 

One object, J0241$-$0326, displays a strong polarization at the $Z$-band and is the most variable object in the sample with polarimetric variability at short and long time scales. Other field dwarfs also show variability in their polarimetric properties indicative of atmospheric patterns that change with rotation and at scales signficantly longer than rotation.  

The comparison of the sample of young field dwarfs with the sample of high-gravity dwarfs from the literature yields that both sets statistically have no obvious difference in their polarimetric intensities, suggesting that the impact of age (atmospheric pressure or gravity) is not detectable at the present level of accuracy of our data. Similarly, there is no clear trend of the linear polarimetry degree with the spectral type in any of the samples. More accurate measurements would be required to study whether such differences and trends may exist.

\section*{Acknowledgements}

Based on observations made with the The William Herschel (WHT) and Nordic Optical Telescopes (NOT), which are operated by the Isaac Newton Group and the Nordic Optical Telescope Scientific Association, respectively, at the Spanish Observatorio del Roque de los Muchachos of the Instituto de Astrof\'\i sica de Canarias in the island of La Palma, Spain. This work has also used observations carried out using the Very Large Telescope at the ESO Paranal Observatory under Program ID: 089.C-0318(A). This research has made use of NASA's Astrophysics Data System and the SIMBAD database, this last one being operated at CDS, Strasbourg, France. Also, this publication makes use of data products from the Wide-field Infrared Survey Explorer, which is a joint project of the University of California, Los Angeles, and the Jet Propulsion Laboratory/California Institute of Technology, funded by the National Aeronautics and Space Administration. This work is partly financed by the Spanish Ministry of Economics and Competitiveness through projects AYA2010-21308-C03-02 and AYA2014-54348-C3-2-R, and ESP2013-48391-C4-2-R and ESP2014-57495-C2-1-R.




\bibliographystyle{mnras}
\bibliography{biblio} 




\appendix
\section{Examples of $q$ and $u$ as a function of aperture}\label{ap1}
\begin{figure*}
\centering
\includegraphics[width=17cm]{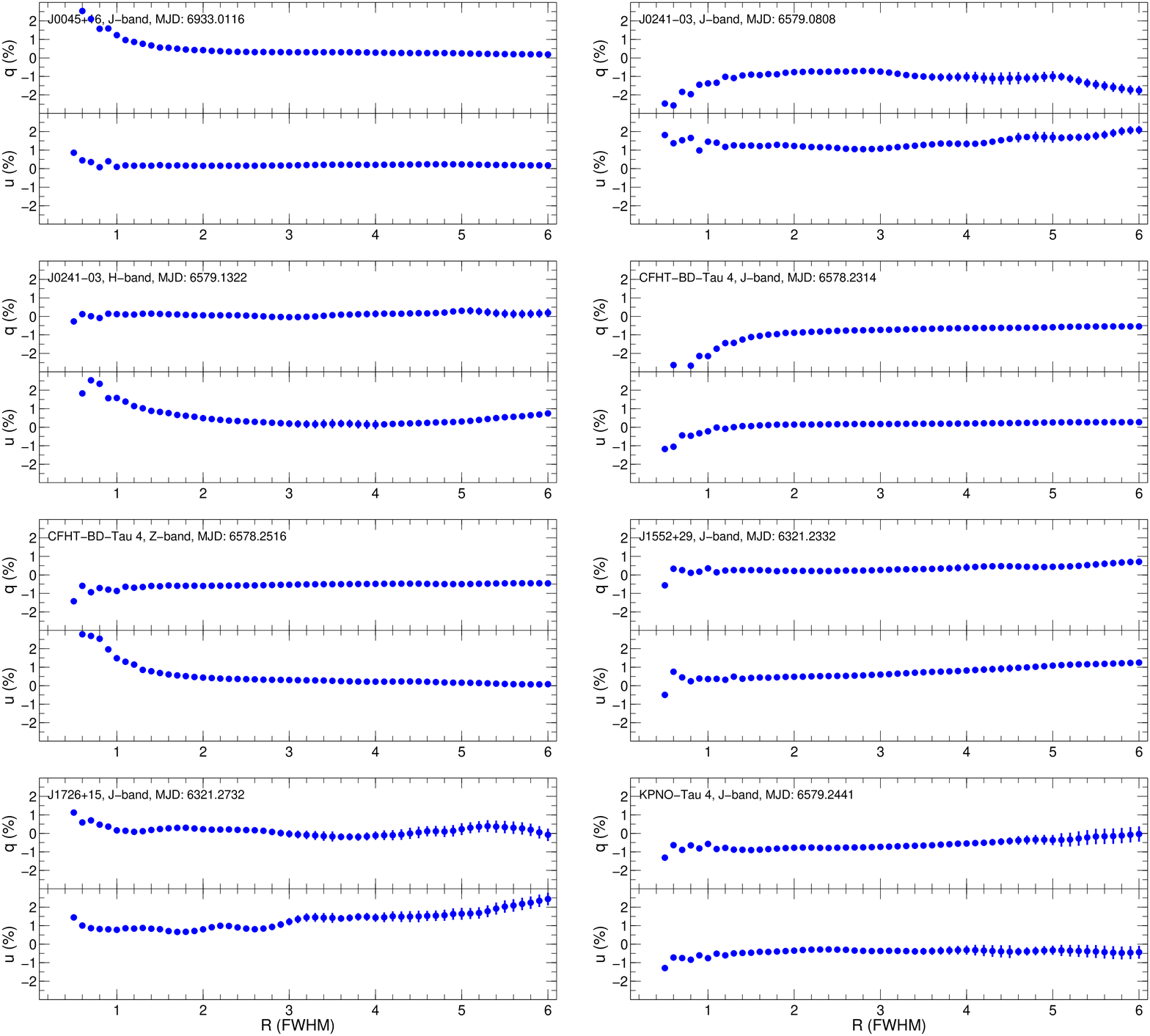}
\caption{Normalized Stokes parameters $q$ and $u$ as a function of the aperture radius (in FWHM unit) for some of our targets. The name of the target, filter, and MJD are indicated in the $q$ panels. For this plot, the sky annulus is fixed at 6 $\times\,$FWHM and 1.5 $\times\,$FWHM in size. Vertical bars indicate the dispersion of the measurements at each aperture over all sky annuli (see Section \ref{polarimetry}); in most cases, these are smaller than the symbol size. We typically selected apertures in the range 2--4 $\times\,$FWHM to compute the mean Stokes parameters.}
\label{figap}
\end{figure*}

In Figure \ref{figap} we plot some examples of the normalized Stokes parameters $q$ and $u$ as a function radii for some of our targets.


\bsp	
\label{lastpage}
\end{document}